\documentclass[preprint,floatfix
,aps,prd,superscriptaddress]{revtex4}

\usepackage[utf8]{inputenc}
\usepackage{graphicx}
\usepackage{tikz}
\usepackage{amssymb,amsmath,mathtools,comment,hyperref}
\usepackage{appendix}

\usepackage{extarrows}
\usepackage{relsize}

\usepackage{bm}
\usepackage{mathrsfs}
\usepackage{xcolor}

\newcommand{\ket}[1]{\left|#1\right\rangle}
\newcommand{\bra}[1]{\left\langle #1 \right|}
\newcommand{\lrp}[1]{\left( #1 \right)}
\newcommand{\lrb}[1]{\left[ #1 \right]}

\newcommand{\rs}[1]{\rho^{\rm #1}}

\begin{document}

\title{Quantum optics meets black hole thermodynamics
 \\
via conformal quantum mechanics:
\\
I. Master equation for acceleration radiation}

\author{A. Azizi}
\affiliation{Institute for Quantum Science and Engineering, Texas A\&M University, College Station, Texas, 77843, USA}
\author{H. E. Camblong}
\affiliation{Department of Physics and Astronomy, University of San Francisco, San Francisco, California 94117-1080, USA}
\author{A. Chakraborty}
\affiliation{Department of Physics, University of Houston, Houston, Texas 77024-5005, USA}
\author{C. R. Ord\'{o}\~{n}ez}
\affiliation{Department of Physics, University of Houston, Houston, Texas 77024-5005, USA}
\affiliation{Department of Physics and Astronomy, Rice University, MS 61, 6100 Main Street, Houston, Texas 77005, USA.}
\author{M. O. Scully}
\affiliation{Institute for Quantum Science and Engineering, Texas A\&M University, College Station, Texas, 77843, USA}
\affiliation{Baylor University, Waco, TX 76706, USA}
\affiliation{Princeton University, Princeton, New Jersey 08544, USA}

\date{\today}
\begin{abstract}
A quantum-optics approach is used to study the nature of the acceleration radiation due to a random atomic cloud falling freely into a generalized Schwarzschild black hole through a Boulware vacuum. 
The properties of this horizon brightened acceleration radiation (HBAR) are analyzed with a master equation that is fully developed in a multimode format.
A scheme for the coarse-graining average for an atomic cloud is considered, with emphasis on the random injection scenario, which is shown to generate a thermal state.
The role played by conformal quantum mechanics (CQM) is shown to be critical for detailed balance via a Boltzmann factor governed by the near-horizon physics, with the unique selection of the Hawking temperature.
The HBAR thermal state is the basis for a thermodynamic framework that parallels black hole thermodynamics.

\end{abstract}
\maketitle

\section{Introduction}

Deep connections between gravitation, quantum theory, and thermodynamics have been uncovered within a consistent framework known as black hole thermodynamics~\cite{hawking76,BH-thermo_reviews}.
This synthesis combines:
 Bekenstein's entropy identification 
with the area of the black hole~\cite{bekenstein1972, bekenstein1973} 
as the basis for 
a generalized second law of thermodynamics (GSL)~\cite{bekenstein1973,bekenstein1974} (along with Hawking's area theorem \cite{christodoulou,hawking71,hawking72}
 and the four laws of black hole mechanics~\cite{bardeen-carter-hawking1973}); and
Hawking's seminal breakthrough \cite{hawking74,hawking75} that black holes can radiate with a definite temperature and associated entropy-area proportionality constant. Hawking radiation and temperature confirm the genuine physical nature of this form of thermodynamics. Related thermal behavior in accelerated systems (including acceleration temperature and radiation) was also identified by Unruh, Davies, and Fulling~\cite{unruh76,fulling76,davies77}.

In this paper, we begin developing the tools needed for a thermodynamic framework that
involves a correspondence between black hole thermodynamics and acceleration radiation.
Most importantly, we identify the conformal nature of the acceleration radiation, using a systematic near-horizon approach, and address its existence and properties for arbitrary initial conditions of an atomic cloud
 freely falling into a Schwarzschild black hole.
Our presentation builds on the insightful quantum optics approach
 developed in 2018~\cite{scully2018}, in which such cloud of atoms emits
 ``horizon brightened acceleration radiation'' (HBAR) with a thermal-like behavior and
 a surprising HBAR area-entropy-flux relation~\cite{scully2018}.
Following Refs.~\cite{camblong2005,nhcamblong-sc,camblong2013,camblong2020,azizi2021}, 
we anticipate that hidden near-horizon conformal symmetry~\cite{gupta01,vaidya00,moretti02}, 
manifests in the leading form of the field equations as
conformal quantum mechanics (CQM) \cite{DFF};
and CQM  governs the predicted thermodynamic behavior.
Specifically, the CQM modes and the corresponding near-horizon geodesics
lead to: (i) the unique Hawking temperature shared by the black hole and the HBAR radiation; (ii) the existence of an HBAR area-entropy-flux relation.
This will be further developed within a larger synthesis in the second article of this series~\cite{HBAR_part-II}.

This article is organized as follows. Section~\ref{sec:setup} consists of a brief outline of the setup of the problem and its component parts. This is followed by 
two parallel tracks: the gravitational concepts described by the background geometry and the statistical concepts described by a quantum-optics approach.
The gravitational logical flow of concepts develops from
 Sec.~\ref{sec:nh_KG_equations} with a derivation of the near-horizon CQM modes and the near-horizon geodesics of the freely falling atoms; and is followed 
  in Sec.~\ref{sec:conformal_steady_state} with a discussion of the conformal aspects of the acceleration radiation, whereby the near-horizon conformal symmetry implies thermal behavior with an associated Hawking temperature in a generalized Schwarzschild background.
The quantum-optics statistical flow of concepts 
begins in Sec.~\ref{sec:master_equation}, where we derive the master equation for the reduced density matrix of the radiation field, generalizing the results of Ref.~\cite{scully2018}; the proof of the thermal nature of the field is expanded with the master equation in Sec.~\ref{sec:density-matrix_steady-state}, where we demonstrate that the CQM framework leads to the thermal field density matrix that completely characterizes the state consisting of all modes of the quantum field. Finally, in Sec.~\ref{sec:conclusion}
we provide concluding remarks and combine the results of the previous sections (with converging parallel tracks) to anticipate the HBAR entropy flux formula of Ref.~\cite{scully2018}---this will be fully developed
in the continuation paper of this series~\cite{HBAR_part-II}.
The appendices include a thorough treatment of the master equation (\ref{app:master_equation_derivation}), the related properties of the average involved in the coarse-grained density matrix (\ref{app:injection-averages_diagonality}), and the geodesics in a generalized Schwarzschild background (\ref{app:geodesics-Schwarzschild}).

\section{Setup of the problem: Generalized Schwarzschild geometry
and acceleration radiation genesis}\label{sec:setup}
The physical setup of the problem involves three interacting systems: a black hole, a quantum field, and an atomic cloud, within an approach that follows the general strategy of Refs.~\cite{scully2018,camblong2020,azizi2021}. 
The three interacting systems are described in terms of a dipole coupling of a scalar field $\Phi$ with a freely falling atomic cloud of two-level systems (``atoms''), within the gravitational background of a black hole. 
In this paper, our main goal is to show that the generation of acceleration radiation of particles by free fall into a black hole is associated with a thermal state, in such a way that {\it the physics of these processes is governed by near-horizon CQM.\/} Conformal symmetry will be highlighted in this work. 
Incidentally, we will use the term ``photon'' to refer to the field quantum; the basic physics of radiation generation and related thermodynamics does not depend on the nature of the field, but we will consider a scalar field for simplicity.

Our derivation in this paper will assume the geometry of a generalized Schwarzschild black hole
in $D$ spacetime dimensions, which is given by
\begin{equation}
ds^{2}=- f (r) \,  dt^{2}+\left[ f(r) \right]^{-1} \, dr^{2}+ r^{2} \, d \Omega^{2}_{(D-2)}\; ,
\label{eq:RN_metric}
\end{equation}
where the last term gives the metric of the $(D-2)$-dimensional sphere that foliates the spacetime. This class of metrics includes $D$-dimensional generalizations of the Schwarzschild metric, of the Reissner Nordstr\"{o}m metric, and of extensions of these with a cosmological constant, and black hole solutions with additional charges. This class of geometries suffices to show how the HBAR-black-hole correspondence arises, and further generalization to non-extremal Kerr black holes will be reported elsewhere~\cite{HBAR_part-II}.

In this geometric background, the physical setup includes a scalar field $\Phi$ and an atomic cloud of two-level systems injected randomly and undergoing free fall towards the generalized Schwarzschild black hole through the Boulware vacuum.
We begin by quantizing the scalar field with
\begin{equation}
    \Phi(
   \mathbf{r}
    ,t) = \sum_{\boldsymbol{s}} \left[
    a_{\boldsymbol{s}} \phi_{\boldsymbol{s}} (\mathbf{r}, t) + \mathrm{h.c.} 
    \right]
    \; ,
        \label{eq:field_expansion}
        \end{equation}
where $\mathrm{h.c.}$ stands for the Hermitian conjugate (adjoint);
and $\mathbf{r}$ denotes the spatial coordinates, i.e.,
$\mathbf{r}= (r, \Omega)$, with $\Omega$ being the angular coordinates,
for the metric~(\ref{eq:RN_metric}).
In Eq.~(\ref{eq:field_expansion}), the field modes $\phi_{\boldsymbol{s}} $ are associated with 
the Boulware vacuum, which is annihilated by the lowering operator
$a_{\boldsymbol{s}}$. In addition,
 the symbol ${\boldsymbol{s}}$ stands for the set of ``quantum numbers'' that provide complete characterization of the mode: 
it includes the frequency $\omega$ of the mode and any additional numbers associated with the geometry and separation of variables. It should be noted that, for finite-box quantization, the frequencies involve a third discrete number. 
 For the Schwarzschild geometry~(\ref {eq:RN_metric}), this set is
${\boldsymbol{s}} =  \{\omega,l,\boldsymbol{m}\}$, where 
$\{l,\boldsymbol{m}\}$ is the set of generalized angular momentum quantum numbers associated with the $(D-2)$-dimensional sphere.

As the atoms fall, they interact with the scalar field via the dipole interaction given by
\begin{equation}
V_I(\tau) = 
g \, \Phi \, \sigma
=
g \sum_{\boldsymbol{s}}
\left[a_{\boldsymbol{s}} \phi_{\boldsymbol{s}}(\mathbf{r} (\tau),t(\tau)) + \mathrm{h.c.} \right]
\lrp{\sigma_- e^{-i\nu \tau} + \mathrm{h.c.}} 
\; ,
\label{eq:QO_interaction_potential}
\end{equation}
where $g$ is the atom-field coupling strength, and  $\sigma_-$ is the atomic lowering operator.
The operator $\sigma = \sigma_- e^{-i\nu \tau} + \sigma_+ e^{i\nu \tau}$ describes the transitions between the two states of the atom. The field modes $\phi_{\boldsymbol{s}}(\mathbf{r} (\tau),t(\tau))$ will be evaluated in the near-horizon region in Sec.~\ref{sec:nh_KG_equations}, as needed in Eq.~(\ref{eq:QO_interaction_potential}), which requires their values parametrized with the proper time of the geodesic trajectories of the atoms. 

In Eq.~(\ref{eq:QO_interaction_potential}), the coupling strength $g$ of the field with the atoms will be sufficiently weak to allow use of perturbation theory for the emission and absorption rates of the atoms as they fall freely through the vacuum. 
Due to the relative acceleration of the field modes with respect to the atoms, a radiation field is emitted by the atoms, despite their locally inertial motion \cite{wilson18,scullyreview}. 
The interaction of the atoms with the radiation field involves processes whereby they will go to an excited state with the emission of a scalar photon.
Up to first order in perturbation theory, the emission probability $P_{{\mathrm e}, {\boldsymbol{s}} }$ for the field mode ${\boldsymbol{s}}$ is given by
\begin{equation}
P_{{\mathrm e}, {\boldsymbol{s}} } 
= \left| \int d\tau \;\bra{1_{\boldsymbol{s}},a}V_I(\tau)\ket{0,b}\right|^2 \equiv g^2 
| I_{{\mathrm e}, {\boldsymbol{s}} } |^2 
\label{eq:P_ex_expression}
\; ,
\end{equation}
where $\ket{b}$ and $\ket{a}$ are the ground and excited states of the atom respectively and 
$-i g I_{{\mathrm e}, {\boldsymbol{s}} }$ is the corresponding probability amplitude.
 Similarly, the absorption probability is given by
\begin{equation}
    P_{{\mathrm a}, {\boldsymbol{s}} }
     = \left|\int d\tau \;\bra{0,a}V_I(\tau)\ket{1_{\boldsymbol{s}},b}\right|^2 \equiv g^2 
     | I_{{\mathrm a}, {\boldsymbol{s}} } |^2 
     \label{eq:P_ab_expression}
     \; ,
\end{equation}
 where $-i g I_{{\mathrm a}, {\boldsymbol{s}} }$ is the absorption probability amplitude. 
 In regards to the dipole potential~(\ref{eq:QO_interaction_potential}), the atomic lowering and raising operators are
 $\sigma_- = \ket{b}\bra{a}$ and $\sigma_+ = \ket{a}\bra{b}$ respectively. 
 As usual, the field states are labeled by their occupation numbers $n_{\boldsymbol{s}}$; here the ground state $\ket{0}$
 and the state $\ket{1_{\boldsymbol{s}}}$ with one photon in mode ${\boldsymbol{s}}$ are considered.
 Thus, we get
\begin{subequations}
    \begin{equation}
        P_{{\mathrm e}, {\boldsymbol{s}} }
         = g^2 \left|\int\; d\tau\; \phi_{\boldsymbol{s}}^*(\mathbf{r} (\tau),t(\tau)) \, e^{i\nu\tau}\right|^2
        \; , 
        \label{eq:P_ex_explicit}
    \end{equation}
    \begin{equation}
        P_{{\mathrm a}, {\boldsymbol{s}} }
        = g^2 \left|\int\; d\tau\; \phi_{\boldsymbol{s}}(\mathbf{r} (\tau),t(\tau)) \, e^{i\nu\tau}\right|^2
        \; 
        . \label{eq:P_ab_explicit} 
    \end{equation}
    \label{eq:P_explicit}%
\end{subequations}
The excitation and absorption probabilities above are critical in finding the field configuration that emerges from the falling atomic cloud. This configuration can be described by the steady-state density matrix of the field, which is derived as the partial trace of the density matrix of the field-atom system (by tracing out the atomic degrees of freedom). As we will see in Secs.~\ref{sec:density-matrix_steady-state} and \ref{sec:conclusion}, the density matrix of the field can be used to 
characterize the thermal properties of the HBAR field and to find thermodynamic relations, including 
the entropy flux due to the photon generation by the excitation of the freely falling atoms.

\section{Basic theory of the master equation for the field modes}\label{sec:master_equation}
With the setup described in Sec.~\ref{sec:setup}, our goal is to derive the master equation for the evolution of the quantum field, an essential ingredient to determine the thermodynamic properties, including the entropy. Generalizing Refs.~\cite{scully2003} and \cite{belyanin2006}, this will be achieved by considering the rate of change of the reduced density matrix ($\rs{\mathcal P}$) for all degrees of freedom of the field, due to the random injection of atoms that fall freely through the Boulware vacuum. This procedure will require a detailed analysis that involves a statistical averaging as well as consideration of all field modes $\phi_{\boldsymbol{s}}$. As pointed out in Sec.~\ref{sec:setup}, the choice of a scalar field is made to simplify the calculations, and the result of the radiation generation is the emission of scalar photons. Thus, we will refer to the field as the photon system (labeled with ${\mathcal P}$), which interacts with the atom (labeled with ${\mathcal A}$).

A complete description of the time evolution of the combined atom-field system is governed by the von Neumann equation satisfied by the density matrix $\rs{{\mathcal P}{\mathcal A}}$.
 In the interaction picture, and up to second order, this equation is
\begin{equation}
\!    \rs{{\mathcal P}{\mathcal A}}(\tau) 
    =  \rs{{\mathcal P}{\mathcal A}}(\tau_0) 
    - i \! \int_{\tau_0}^{\tau} d\tau' [V_I(\tau'),\rs{{\mathcal P}{\mathcal A}}(\tau_0)] 
    - \! \int_{\tau_0}^{\tau} \! \! \! d\tau' \int_{\tau_0}^{\tau'} \! \! \! d\tau'' [V_I(\tau'),[V_I(\tau''),\rs{{\mathcal P}{\mathcal A}}(\tau_0)]] 
    \label{eq:density-matrix_evolution}
    \, .
\end{equation}
The initial state of the combined system is the tensor product $\rs{{\mathcal P}{\mathcal A}}(\tau_0) = \rs{\mathcal P}(\tau_0) \otimes \rs{\mathcal A} (\tau_0)$, as the atoms and field are initially uncorrelated. The time parameter $\tau$ to be used a priori for our problem in Eq.~(\ref{eq:density-matrix_evolution}) is the proper time of free-fall trajectories, as this describes the dynamics of the atoms that generate the radiation.

We will consider a cloud of atoms that acts as a reservoir. The atoms are injected in their ground state, so that an atom's initial density matrix (at time $\tau_0$) is $ \rs{\mathcal A} = \ket{b}\bra{b}$. The evolution of the system of interest, the radiation field, will be properly averaged over a distribution of injection times. 
While the density matrix $\rs{\mathcal A} $ is defined at time $\tau_{0}$, we will assume a Markovian property for the system's dynamics, whereby the predictions for the evolution of the field density matrix can be made by using the reservoir's initial state alone. This is a memorylessness property of the reservoir, which means that its memory is short, i.e., its density matrix does not change appreciably over a significant evolution of the field system.
 
As a result, the effective evolution of the field is obtained by a twofold process:
tracing over the atomic degrees of freedom (with $\mathrm{Tr_{\mathcal A}}$) to get the reduced field density matrix $\rs{\mathcal P}  =  \mathrm{Tr_{\mathcal A}} \, \left(  \rs{{\mathcal P}{\mathcal A}} \right)$; and performing an appropriate averaging procedure over a time scale larger than the reservoir's memory time---here, this is characteristic time scale that captures a representative distribution of injection times. This approximate evolution, governed by a master equation~\cite{scullybook,meystrebook}, is said to be coarse-grained, with the implication that the effective field states are described by an approximate coarse-grained reduced density matrix $\rs{\mathcal P}$.  

The averaging procedure that leads to the coarse-grained density matrix can be established as follows---and for this problem, we will loosely refer to it as ``injection average.'' 
This is a standard procedure in quantum optics, where the experimental setup involves an optical cavity.
The cavity analog for the generalized Schwarzschild geometry corresponds to a spatial region bounded by constant values of the Schwarzschild spatial coordinates, with the experimental devices being carried by a set of static observers, and such that the times are specified via the coordinate time $t$.
Consider one atom labeled with subscript $a$, which is injected at an initial time $t_{0} =t_{i,a}$ as measured in the cavity.
The atom will undergo geodesic motion with the equations given in Appendix~\ref{app:geodesics-Schwarzschild}. As a result, there exists a relationship $\tau = \tau(t)$ specifically given by Eq.~(\ref{eq:tau-from-t}); therefore, the time parameter $\tau$ can be replaced by $t$, and  
the corresponding exact or ``microscopic'' change in the field density matrix due to the injection of the single atom is $\delta  \rs{\mathcal P}_{a} \equiv \delta \rs{\mathcal P} (t; t_{i,a}) 
  =  \rs{\mathcal P}(t) - \rs{\mathcal P}(t_{i,a})$. (Here, there is an abuse of notation, $\tau(t) \rightarrow t$.)
  Then, for a cloud or ensemble of atoms, the course-grained (``macroscopic'') change is $\Delta  \rs{\mathcal P} = \sum_{a} \delta \rho_{a}$, whence the coarse-grained rate of change, for which we will use the overdot notation, is~\cite{scullybook}
\begin{equation}
\dot{\rho}^{\mathcal P} \equiv \frac{\Delta \rs{\mathcal P}}{\Delta t} = \mathfrak{r}  \, \frac{1}{\Delta N} 
\, \sum_{a} \delta \rs{\mathcal P} (t; t_{i,a}) 
=  \mathfrak{r} \,  \overline{\delta \rs{\mathcal P} }
\; ,
\label{eq:dot-rho_coarse-grained}
\end{equation}
where $\mathfrak{r}= \Delta N/\Delta t $ is the atom injection rate and $\overline{\delta \rs{\mathcal P}}$ stands for the average microscopic change with respect to particle injection. 
The rate of change in Eq.~(\ref{eq:dot-rho_coarse-grained}) is computed with respect to the generalized Schwarzschild coordinate time $t$.
Indeed, as the coarse-grained density matrix is used to describe states of the field ${\mathcal P}$ in the Boulware vacuum, and not of the atoms, its natural (coarse-grained) evolution should be governed by this coordinate time $t$.
In the quantum optics analogy, the coordinate $t$ corresponds to times experienced at given locations of an optical cavity.
 This is in contradistinction with the evolution of the complete density matrix 
$  \rs{{\mathcal P}{\mathcal A}}(\tau) $
as in Eq.~(\ref{eq:density-matrix_evolution}),
which is governed by the proper time $\tau$.

The injection average can be performed with respect to the initial-time variable $t_{i,a}$ that governs the statistical distribution of the cloud as reservoir. 
This choice of statistical average corresponds to the following operational procedure
for the formation of the atomic cloud:
(i) the atoms are released with given initial conditions from a specified radial-coordinate value $r_{i}$ at variable Schwarzschild coordinate times $t_{i,a}$;
 (ii) the atoms are allowed to follow ingoing geodesics towards the event horizon.
Each geodesic is described by the proper time parameter, starting with a value $\tau_{i,a}$ and reaching a later point with value $\tau_{f,a}$; because of the static nature of the metric and the specified initial conditions, one can assign uniform values of the proper time along the geodesics with a given fixed value 
$\tau_{i,a} = \tau_{i}$. Details on the geodesics are discussed in Sec.~\ref{sec:nh_KG_equations}.
The times $t_{i,a} $
are the coordinate times for injection events located at $r= r_{i}$ in the Schwarzschild geometry. 
These events can occur at different moments in the evolution of the system; thus, $t_{i,a} $ are arbitrary additive constants in the geodesic relationship $t=t(\tau)$,
i.e., from Eq.~(\ref{eq:tau-from-t}), $t= t_{i,a} + H^{-1}(\tau - \tau_{i})$.
Thus, in the Schwarzschild analog ``cavity,'' one can use the times $t_{i,a}$ as variables that define a statistical distribution, with the other parameters fixed. The corresponding statistical average is defined by considering a number of atoms $\Delta N $ during a time interval $T$, such that 
$(1/\Delta N ) \sum_{a} =  \int d t_{i,a} f (t_{i,a}) $, where $f (\xi)$ is the probability distribution of the random variable. Then,
\begin{equation}
\overline{\delta \rs{\mathcal P}}
=
 \int d t_{i} \, f ( t_{i}) 
\,
  \delta \rs{\mathcal P} (t; t_{i}) 
\; ,
\label{eq:injection-average}
\end{equation}
with the index $a$ removed in favor of a generic initial time $t_{i} \equiv t_{0}$. This procedure can be applied to any other relevant field quantity. For a completely random distribution of injection times, a uniform distribution $ f ( t_{i}) = 1/T$ can be chosen; this choice will be made in Appendix~\ref{app:injection-averages_diagonality}.

 
Therefore, from Eqs.~(\ref{eq:density-matrix_evolution}) and (\ref{eq:dot-rho_coarse-grained}), the coarse-grained rate of change of the reduced field density matrix~\cite{scullybook,meystrebook}, when averaged over a distribution of injection times~(\ref{eq:injection-average}), is given by
\begin{equation}
\dot{\rho}^{\mathcal P}
 = -\mathfrak{r} \, \int_{\tau_i}^{\tau_f= \tau} \int_{\tau_i}^{\tau'} d\tau'd\tau''\,\, \mathrm{Tr_{\mathcal A}}\left[V(\tau'),\left[V(\tau''),\rs{\mathcal P}(t (\tau_i) )\otimes\rs{\mathcal A}(\tau_i)\right]\right]  \label{eq:masterequation1}
 \; ,
\end{equation}
where the lower limit $\tau_i$ is the initial (injection) time and the upper limit $\tau_f$ is the final (observation) time, equal to the effective time that defines the argument of the density matrix on the left-hand side of the equation. In the quantum-optics analogy, these are the entry and exit times of the atoms in the cavity. The leading second-order of perturbation theory in Eq.~(\ref{eq:masterequation1}) is due to the fact 
that the interaction potential $V_{I}$ of Eq.~(\ref{eq:QO_interaction_potential}) contains only one raising or lowering operator in $\sigma$, so that the first-order term is vanishing, as it is proportional to 
$\mathrm{Tr_{\mathcal A}} \left( \sigma  \ket{b}\bra{b} \right) =  \bra{b} \sigma \ket{b}=0$.
 Moreover, when the integral on the right-hand side of Eq.~(\ref{eq:masterequation1}) is rewritten using Eq.~(\ref{eq:QO_interaction_potential}), 
    \begin{equation}
  \! \! \!   \int_{\tau_i}^{\tau_f = \tau} \int_{\tau_i}^{\tau'}
   \! \! d\tau'd\tau''\,\,\mathrm{Tr_{\mathcal A}} [V', [V'', \rs{{\mathcal P}{\mathcal A}}  ] ] 
    =
     g^2  \! \!  \int_{\tau_i}^{\tau_f = \tau} \int_{\tau_i}^{\tau'}  \! \! \! 
     d\tau'd\tau''
    \,
     \mathrm{Tr_{\mathcal A}}
    \left[\Phi' \sigma' , [ \Phi'' \sigma '' , \rs{\mathcal P} \otimes \rs{\mathcal A} ] \right] 
    \, ,
    \label{eq:int-trace-commutator}
\end{equation}
where the shorthand notation $V'$ stands for $V(\tau')$, and $V''$ for $V(\tau'')$; and similarly for the operators $\Phi$ and $\sigma$. Two key steps are needed to untangle this expression. 
First, the double commutator can be regrouped in pairs of adjoint operators, 
\begin{equation}
     \left[V',[V'',\rs{{\mathcal P}{\mathcal A}}] \right] 
     = V' [V'',\rs{{\mathcal P}{\mathcal A}}] + \mathrm{h.c.} 
   = \left( V'V''\rs{{\mathcal P}{\mathcal A}} + \mathrm{h.c.}  \right)- \left(V'\rs{{\mathcal P}{\mathcal A}} V'' + \mathrm{h.c.} \right)
     \label{eq:commutator-VVrho}
     \; ,
\end{equation}
 due to the Hermiticity of the potential and the density matrix. This pattern arises because the commutator of two Hermitian operators is anti-Hermitian, but the double commutator is Hermitian. The same regrouping identity can be applied to $\left[\Phi',[\Phi'',\rs{{\mathcal P}{\mathcal A}}] \right]$ after taking the partial trace $\mathrm{Tr_{\mathcal A}}$.
Second, taking into account the operator ordering, the partial trace yields
\begin{equation}
  \mathrm{Tr_{\mathcal A}} 
 \left[ 
  \pi \left(    
     \sigma' \sigma'' \rs{\mathcal A}\right)
      \right] 
          =
     e^{- i \, \mathrm{sgn} ( \pi) \, \nu \tau'} e^{ i \, \mathrm{sgn} (\pi ) \, \nu \tau''}
     \label{eq:atomic-trace}
\end{equation}
where $\pi$ is any permutation of the operators, with signature $\mathrm{sgn} (\pi ) $ accounting for the cyclic property of the trace.
Then, with the steps and substitutions shown in Appendix~\ref{app:master_equation_derivation},
the field master equation~(\ref{eq:masterequation1}) can be recast in a more explicit form directly in terms of the field operators $\Phi$,
 \begin{equation}
  \dot{\rho}^{\mathcal P}
 = -\mathfrak{r} g^2 \,
      \left\{
\left( \int\displaylimits_{I} 
\Phi' \Phi'' \, {\rho}
 + \int\displaylimits_{II} 
{\rho} \,  \Phi'  \Phi'' 
\right)
d^2 \tau \,    e^{- i  \nu  \tau'} e^{ i \nu \tau''}
 -
\int\displaylimits_{I+II}
d^2 \tau  \,    e^{ i   \nu \tau'} e^{ - i \nu \tau''}
\Phi' \, {\rho} \, \Phi'' 
    \right\}
    \label{eq:masterequation2}
    \; .
\end{equation}
The double integrals on the right-hand side of Eq.~(\ref{eq:masterequation2}) are performed with $d^2 \tau = d \tau' d \tau''$ in regions $I \equiv \tau' > \tau''$ and $II \equiv \tau' < \tau''$; this rearrangement corresponds to the exchange of symbols $\tau' \leftrightarrow \tau'$ in the adjoint expressions involved in the commutators. This is discussed in Appendix~\ref{app:master_equation_derivation}, where we also show that further expansion of the operators $\Phi$ on the right-hand side of Eq.~(\ref{eq:masterequation2}), in terms of field creation and annihilation operators, gives a more transparent and useful form of the master equation. 

In what follows, choosing a specific atomic-cloud distribution of injection times, the resulting density matrix elements to be used in Eqs.~(\ref{eq:masterequation1}) and (\ref{eq:masterequation2}) will be denoted by 
$\rs{\mathcal P}_{n,m} = \bra{n} {\rho} \ket{m}$. To simplify the notation, for the remainder of the paper, we will use ${\rho}$ to denote the density matrix of the field (dropping the superscript $\rm {\mathcal P}$). 
Moreover, in Appendix~\ref{app:injection-averages_diagonality}, we justify the statement that, when the {\em atomic-cloud distribution of injection times is random\/}, all the off-diagonal matrix elements average out to zero. Thus, only the diagonal elements $\rs{\mathcal P}_{n,n}$ are nonvanishing for a randomly injected cloud; this is due to the removal of the random phase associated with the injection times, which is only effective for terms involving pairs of creation and annihilation operators of the same kind. 
Thus, for the single-mode version of the reduced field density matrix and random injection times, the master equation becomes
\begin{equation}
    \dot{\rho}_{n,n} =
    - R_{{\mathrm e}, {\boldsymbol{s}} } \big[(n+1) \, {\rho}_{n,n} - n \,  {\rho}_{n-1,n-1}\big] -  
    R_{{\mathrm a}, {\boldsymbol{s}} }\big[n \, {\rho}_{n,n}-(n+1) \, {\rho}_{n+1,n+1} \big] 
    \label{eq:master_equation_final_single-mode}
    \; ,
\end{equation}
where $R_{{\mathrm e}, {\boldsymbol{s}} } = \mathfrak{r} \, P_{{\mathrm e}, {\boldsymbol{s}} } = \mathfrak{r} \, g^2 |I_{{\mathrm e}, {\boldsymbol{s}} }|^2$ and 
$R_{{\mathrm a}, {\boldsymbol{s}} }= \mathfrak{r} \, R_{{\mathrm a}, {\boldsymbol{s}} }
 = \mathfrak{r} \, g^2 |I_{{\mathrm a}, {\boldsymbol{s}} }|^2$ are the emission and absorption rate coefficients of the given mode ${\boldsymbol{s}}$, defined in terms of the probability amplitudes as in Eqs.~(\ref{eq:P_ex_expression})--(\ref{eq:P_ab_expression})  and (\ref{eq:P_ex_explicit})--(\ref{eq:P_ab_explicit}). Similarly, the occupation number $n$ is a function of the given field mode, and this will be further explained next. Equation~(\ref{eq:master_equation_final_single-mode}) is the particular type of master equation proposed in Refs.~\cite{scully2018}, \cite{scully2003}, and \cite{belyanin2006}. We will generalize the master equation to describe all the modes of the field simultaneously, as required for the proper computation of additive thermodynamic functions, including the entropy.

While Eq.~(\ref{eq:master_equation_final_single-mode}) captures the essence of the master equation by showing the interplay between states with $n$ and $n \pm 1$ photons, the more general multimode expression derived in
Appendix~\ref{app:master_equation_derivation} provides the foundation for a complete description of the Planck distribution of the scalar photon field studied in this paper. The notation for the multimode framework is more involved, and we will use some conventional choices, with adjustments to simplify the final results, as follows. First, the single-mode quantum numbers $ {\boldsymbol{s}} $ can be chosen in an ordered sequence  
${\boldsymbol{s}}_1,  {\boldsymbol{s}}_2,  \ldots {\boldsymbol{s}}_{j}$ \ldots; and the state of the field in the occupation number representation involves the number 
 $ n_{j} \equiv n_{ {\boldsymbol{s}}_{j} } $ 
of scalar photons in the state ${\boldsymbol{s}}_{j}  $. Thus, with the notation 
$ \boldsymbol{ \left\{  \right. } n  \boldsymbol{\left. \right\}}    \equiv
\boldsymbol{ \left\{  \right.  } n_{1}, n_{2}, \ldots , n_{j } , \ldots \boldsymbol{\left. \right\}  }$, 
the multimode state with given occupation numbers is the tensor product of single-mode states: 
 $
 \left|   \boldsymbol{ \left\{  \right. } n  \boldsymbol{\left. \right\}}  \right\rangle 
 \equiv
\left|  n_{1}, n_{2}, \ldots , n_{j } , \ldots  \right\rangle
$.
Second, the elements of the density matrix are
$ {\rho} \bigl({ \boldsymbol{ \left\{  \right. } n  \boldsymbol{\left. \right\}  };  \boldsymbol{ \left\{  \right. } n' \boldsymbol{\left. \right\}  } } \bigr)
\equiv
\boldsymbol{ \left\langle \right. }
\boldsymbol{ \left\{  \right. }
 n  \boldsymbol{\left. \right\}  } 
\boldsymbol{ \left. \right|}
\rho
\boldsymbol{ \left| \right. }
  \boldsymbol{ \left\{  \right. } n' \boldsymbol{\left. \right\}  } 
\boldsymbol{\left. \right\rangle  }
\equiv
  {\rho}_{ n_1,n_2, \ldots ;   n'_1,n'_2, \ldots } $
  (where the semicolon is used for notational clarity).
Finally, the diagonal elements of the density matrix will be denoted by
  \begin{equation}
  {\rho}_{\rm diag} (  \boldsymbol{ \left\{  \right. } n  \boldsymbol{\left. \right\}  } )
  \equiv
 {\rho} \bigl({ \boldsymbol{ \left\{  \right. } n  \boldsymbol{\left. \right\}  } ; \boldsymbol{ \left\{  \right. } n  \boldsymbol{\left. \right\}  } }\bigr)
\; \; \; \text{or} \; \; \;
 \rho_{\rm diag}    (  n_1, n_2, \ldots ) 
   \equiv
  {\rho}_{  n_1,n_2, \ldots  ;   n_1,n_2, \ldots  }
    \label{eq:diagonal-density_multimode}
    \; .
\end{equation}
Moreover, we will use the shorthand notation 
\begin{equation}
  \boldsymbol{ \left\{  \right. } n  \boldsymbol{\left. \right\}}_{n_j + q}  
  \equiv
 \left\{  \right.  
 n_{1}, n_{2}, \ldots , n_{j }+ q , \ldots \boldsymbol{\left. \right\}  }
 \; ,
 \label{eq:number-shift_multimode}
 \end{equation}
  where $q$ is an integer number,
   to represent the occupation-number set where the mode labeled by $j$ (i.e., ${\boldsymbol{s}}_j$) has a number shift
   $n_{j} \rightarrow n_{j} + q$; that is, given a particular value $n_{j}$ (stated within the given equation), the appropriate value to use for the occupation number is shifted as specified. Furthermore, these shifts can be implemented multiple times within a set (more than one occupation number), and for the different rows and columns of the density matrix.
   
Then, as shown in Appendix~\ref{app:master_equation_derivation}, the multimode master equation for our problem takes the form
\begin{equation}
\begin{aligned}
 \dot{\rho}_{\rm diag}(  \boldsymbol{ \left\{  \right. } n  \boldsymbol{\left. \right\}  } )  
 =
    - 
     \sum_{j}
     &
     \left\{
     R_{{\mathrm e},\, j}  \big[(n_j+1) \,
   {\rho}_{\rm diag} (  \boldsymbol{ \left\{  \right. } n  \boldsymbol{\left. \right\}  } )
      - n_j \,
   {\rho}_{\rm diag} (  \boldsymbol{ \left\{  \right. } n  \boldsymbol{\left. \right\}  }_{n_j -1} )
            \big] \right.
          \\
           &
            \left.
            +  
        R_{{\mathrm a},\, j} \big[ n_j \,
    {\rho}_{\rm diag} (  \boldsymbol{ \left\{  \right. } n  \boldsymbol{\left. \right\}  } )
      - (n_j + 1)  \,
     {\rho}_{\rm diag} (  \boldsymbol{ \left\{  \right. } n  \boldsymbol{\left. \right\}  }_{n_j +1} )
                           \big] \right\}
    \label{eq:master_equation_final_multimode}
    \; ,
\end{aligned}
\end{equation}
under the assumption that only the diagonal elements are relevant.
In Eq.~(\ref{eq:master_equation_final_multimode}), again,
\begin{equation}
\begin{aligned}
& R_{{\mathrm e}, j } = \mathfrak{r} \, P_{{\mathrm e}, j } = \mathfrak{r} \, g^2 |I_{{\mathrm e}, j }|^2
\\
& R_{{\mathrm a}, j}= \mathfrak{r} \, R_{{\mathrm a}, j} = \mathfrak{r} \, g^2 |I_{{\mathrm a}, j}|^2
\end{aligned}
\label{eq:emission-absorption_coeffs}
\end{equation}
 are the emission and absorption rate coefficients of the given mode ${\boldsymbol{s}}_{j}$. As stated above, the diagonal-reduction property is achieved, as shown in Appendix~\ref{app:injection-averages_diagonality}, 
  for {\em random injection times\/}---otherwise additional terms would appear on the right-hand side of Eq.~(\ref{eq:master_equation_final_multimode}), corresponding to transitions with jumps in occupation number by two units as well as involving different modes. The functional form of Eq.~(\ref{eq:master_equation_final_multimode}) only includes changes of one mode at a time, i.e., the modes are effectively independent. This mode independence is guaranteed by enforcing two distinct conditions: the field interacts with the atoms and the gravitational background, but does not self-interact; and the atom-injection distribution is random.

\begin{figure}[h]
    \centering
    \includegraphics[width=0.95\linewidth]{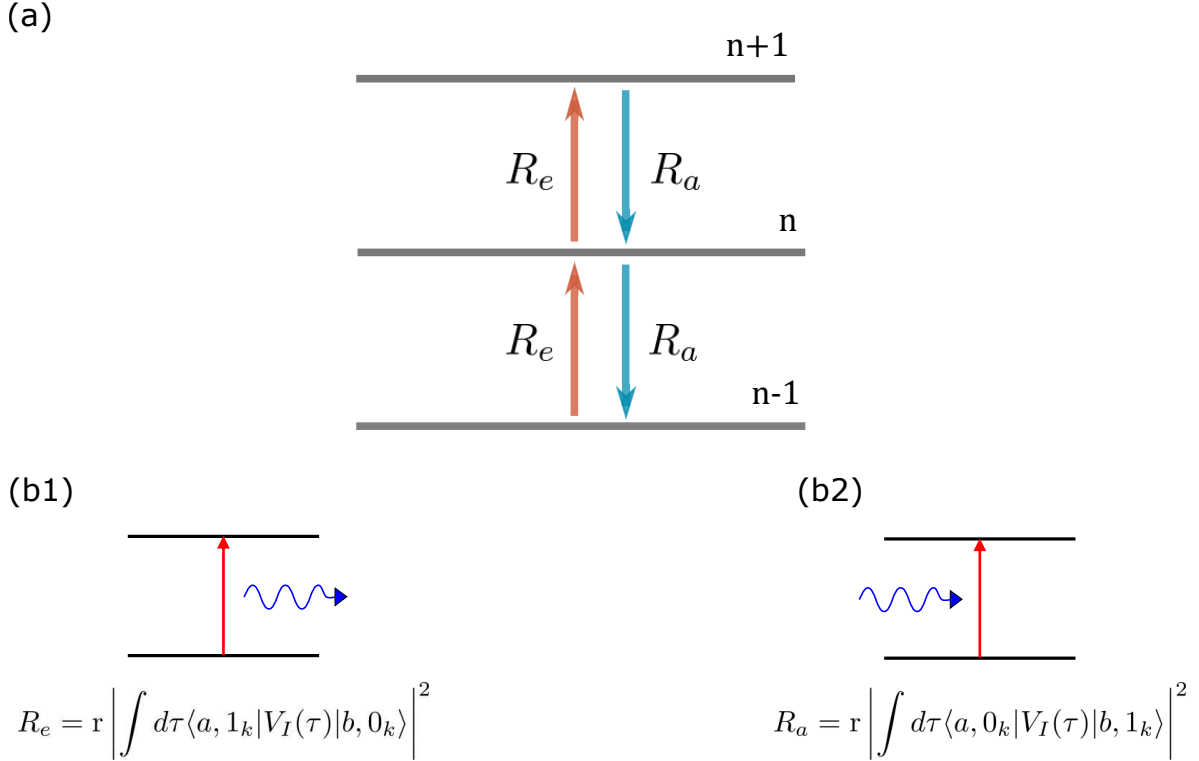}
    \caption{Probability flow described by the master equation~(\ref{eq:master_equation_final_multimode}). In (a) The three levels of the field mode is shown with the transition probabilities indicated. (b1) and (b2) show the atomic transitions corresponding to the emission and absorption rates.  
    Detailed balance of absorption and emission establishes a steady state analyzed 
    in Sec.~\ref{sec:density-matrix_steady-state}.}    
    \label{fig:detailed-balance}
\end{figure}

The physical meaning of Eq.~(\ref{eq:master_equation_final_multimode}) can be described in terms of the 
photon distribution functions
$p(  \boldsymbol{ \left\{  \right. } n  \boldsymbol{\left. \right\}  } )
\equiv
{\rho}_{\rm diag}(  \boldsymbol{ \left\{  \right. } n  \boldsymbol{\left. \right\}  } )$, i.e., 
the probabilities of states  $\boldsymbol{ \left\{  \right. } n  \boldsymbol{\left. \right\}}$
with given numbers of
 photons in the field. This amounts to a flow of probability~\cite{scullybook} into and out of the 
$
  \boldsymbol{ \left\{  \right. } n  \boldsymbol{\left. \right\}}
$
  state, 
from and into the neighboring states
$
  \boldsymbol{ \left\{  \right. } n  \boldsymbol{\left. \right\}}_{n_j \pm 1}  
  \equiv
 \left\{  \right.  
 n_{1}, n_{2}, \ldots , n_{j } \pm 1 , \ldots \boldsymbol{\left. \right\}  }$,
 for a particular mode ${\boldsymbol{s}}_{j}$, as shown in Fig.~\ref{fig:detailed-balance}.
 When the master equation takes this specific diagonal form, it has the same properties as a Planck distribution, as we will show in Sec.~\ref{sec:density-matrix_steady-state}, where we will provide a complete characterization of the ensuing thermal state. Before proceeding with such detailed analysis 
 of the consequences of Eq.~(\ref{eq:master_equation_final_multimode}),
 we have to determine the emission and absorption probabilities~(\ref{eq:P_ex_explicit})
 and (\ref{eq:P_ab_explicit}),
 using the interaction Hamiltonian~(\ref{eq:QO_interaction_potential}).
 This program will first require a careful examination of the field equations, near-horizon CQM, and the spacetime trajectories of the atoms, as will be discussed in Sec.~\ref{sec:nh_KG_equations}. 

\section{Near-horizon physics and conformal quantum mechanics in generalized Schwarzschild geometry: field equations and geodesics}
\label{sec:nh_KG_equations}
The field modes that form a basis for the quantization of the scalar field are solutions to the Klein-Gordon  equation in curved spacetime,
\begin{equation}
\left( \Box - \mu_\Phi^{2}  \right) \Phi \equiv \frac{1}{ \sqrt{-g} }\partial_{\mu} \left(\sqrt{-g} \,g^{\mu \nu}\,\partial_{\nu} \Phi\right)- \mu_\Phi^{2} \Phi= 0\; ,
\label{eq:Klein_Gordon_basic}
\end{equation}
where $\mu_\Phi$ is the mass of the scalar field. When the specific form of the generalized Schwarzschild metric~(\ref{eq:RN_metric}) is used, Eq.~(\ref{eq:Klein_Gordon_basic}) defines a differential equation for the mode functions valid at all distances outside the event horizon. In order to see the role played by the near-horizon region, we will obtain a simplified equation for the field that highlights the governing role played by conformal symmetry in black hole thermodynamics. In this section, following Refs.~\cite{camblong2005,nhcamblong-sc}, we will briefly discuss the near-horizon reduction process and the functional form of the field modes.

The quantization of the scalar field is enforced by Eq.~(\ref{eq:field_expansion}), with the procedure and notation discussed in Sec.~\ref{sec:setup}.
This is the standard approach for quantum fields in curved spacetime~\cite{Birrell-Davies}, which requires
finding the basis modes for the fields.
Because of its time-translation invariance, the generalized Schwarzschild metric~(\ref{eq:RN_metric}) admits 
a Killing vector $  \boldsymbol\xi = \partial_{t}$, with respect to which we can define positive frequency modes with frequency 
$\omega$; this Killing vector is timelike all the way to the event horizon, where it is null. 
In addition, the spherical symmetry of the metric~(\ref{eq:RN_metric}) allows for 
 the separation of variables 
\begin{equation}
    \phi_{\boldsymbol{s}} (r,\Omega, t) = R(r) Y_{l \boldsymbol{m}} (\Omega)
e^{-i\omega t} 
       \; ,
          \label{eq:KG_separation}
\end{equation}
where $Y_{l \boldsymbol{m}} (\Omega)$
are the hyperspherical harmonics as solutions of the Laplacian on the $(D-2)$-sphere.

We now examine the behavior of the Klein-Gordon equation near the horizon by defining a variable $x = r - r_+$, where $r_+$ is the radius of the outer event horizon, defined by the largest root of the equation $f(r) = 0$. To extract the near-horizon behavior, we use a systematic Taylor expansion of the radial equation around $r_+$. Furthermore, we also apply a Liouville transformation of the form $R(x) \propto x^{-1/2} u(x)$.
This procedure was shown in detail in 
 Refs.~\cite{camblong2020,camblong2005}.
 Then, the radial part of the scalar field near the horizon,
keeping terms up to the leading order in $x$, reduces to the differential equation
\begin{equation}
    u''(x) + \frac{\lambda}{x^2}\left[1+\mathcal{O}(x)\right] = 0
    \; , \label{eq:CQM_Hamiltonian}
\end{equation}
where 
\begin{equation}
\lambda = \frac{1}{4} + \Theta^2\, , \hskip 4em \Theta = \frac{{\omega}}{2\kappa} 
\; ,
\end{equation}
in which $\kappa$ is the surface gravity of the black hole defined by $\kappa = -(\nabla_\mu \xi_\nu) (\nabla^\mu \xi^\nu)/2 = f_{+}'/2 $. We have also used the short-hand notation $f_+'\equiv f'(r_+)$, where prime denotes derivative with respect to $r$;  in this paper, $f_+' \neq 0$, since we are considering non-extremal  black holes.

The symmetry of properties of Eq.~(\ref{eq:CQM_Hamiltonian}) are
crucial for an appropriate interpretation of the governing near-horizon physics. This is a Schr\"{o}dinger-like equation with Hamiltonian $\mathscr{H} = p_x^2/2 + V_{\rm eff}(x)$, where $V_{\rm eff}(x) = - \lambda/x^2$ is the effective potential term. This Hamiltonian $\mathscr{H}$ is classically scale invariant, with an enlarged SO(2,1) symmetry group that describes a $(0+1)$-dimensional conformal field theory, known as conformal quantum mechanics (CQM). The algebra of this SO(2,1) group is generated by  $\mathscr{H} $, combined with the dilation operator $D$ and the special conformal operator $K$. In this article, we will not explore any further the group-theoretic structure of the problem, but will mostly rely on the consequences of its scale symmetry.

A pair of linearly independent solutions of Eq.~(\ref{eq:CQM_Hamiltonian}) is given by $u(x) = x^{1/2\pm i\Theta}$. These are oscillatory functions with a logarithmic phase that is the signature of scale invariance.
Combined with their time dependence, these solutions give outgoing and ingoing CQM modes, 
\begin{equation}
    \Phi^{\pm (CQM)}_{{\boldsymbol{s}} } 
     \stackrel{(\mathcal H)}{\propto}
     x^{\pm i\Theta} Y_{lm}(\Omega) e^{-i\omega t} 
    \; ,
    \label{eq:Schwarzschild_CQM_modes}
\end{equation}
where $ \stackrel{(\mathcal H)}{\propto}$ (as well as $ \stackrel{(\mathcal H)}{\sim}$ below)
denotes the hierarchical near-horizon expansion. 

We will use the CQM modes of Eq.~(\ref{eq:Schwarzschild_CQM_modes}) in Sec.~\ref{sec:conformal_steady_state} to find the emission and absorption rates of the free-falling atoms. For this purpose, we also need the near-horizon geodesic equations for the class of generalized Schwarzschild metrics~\cite{camblong2020}; see summary in Appendix~\ref{app:geodesics-Schwarzschild}.
In particular, the near-horizon limiting forms 
for the time and radial coordinates satisfy the following functional relationships $\tau=\tau(x)$ and $t=t(x)$:
\begin{align}
    \tau & \stackrel{(\mathcal H)}{\sim}  - k x +\mathrm{ const.} +  \mathcal{O}(x^2)\;,\label{eq:tau_in_x}\\
    t &  \stackrel{(\mathcal H)}{\sim}  -\frac{1}{2 \kappa}\ln x - C \, x
    + \mathrm{ const.} + \mathcal{O}(x^2)
    \; . \label{eq:t_in_x} 
\end{align}
In Eq.~(\ref{eq:t_in_x}), $k=1/e$ and
the constant $C$, governing the linear term in $x$, is given in Eq.~(\ref{eq:constant-C}).
 These constants, as shown in Sec.~\ref{sec:conformal_steady_state}, do not play a direct role in the radiation formula. 

\section{HBAR in generalized Schwarzschild geometry:
Conformal aspects of the acceleration radiation} \label{sec:conformal_steady_state}
With the near-horizon field modes and spacetime trajectory equations, we can now find the emission and absorption rates for the freely falling atoms in the generalized Schwarzschild geometry. This calculation, whose setup is evident from Eqs.~(\ref{eq:P_ex_explicit}) and (\ref{eq:P_ab_explicit}), will be thoroughly worked out next. Our treatment will show that the final results are governed by CQM. In effect, this calculation can be performed by using the purely outgoing CQM modes described in Eq.~(\ref{eq:Schwarzschild_CQM_modes}), normalized as asymptotically exact WKB local waves~\cite{nhcamblong-sc}; therefore,
\begin{subequations}
    \begin{equation}
            P_{{\mathrm e}, {\boldsymbol{s}} } 
            \stackrel{(\mathcal H)}{\sim}\; 
            g^2k^2 \left|\int_0^{x_f} \; dx\; x^{-i\Theta} 
            e^{i{\omega} t(x)} e^{i\nu\tau(x)} \right|^2
        \; , \label{eq:P_ex_CQM}
    \end{equation}
    \begin{equation}
            P_{{\mathrm a}, {\boldsymbol{s}} }  
            \stackrel{(\mathcal H)}{\sim}\; 
            g^2k^2 \left|\int_0^{x_f} \; dx\; x^{i\Theta}  e^{-i{\omega} t(x)} e^{i\nu\tau(x)} \right|^2
        \; , 
        \label{eq:P_ab_CQM} 
    \end{equation}
    \label{eq:P_CQM}%
\end{subequations}
where $k=1/e$ and $x_f$ is an approximate upper limit that demarcates the upper boundary of the region of validity for the near-horizon approximation.
Using Eqs.~(\ref{eq:tau_in_x}) and (\ref{eq:t_in_x}), the emission rate becomes
\begin{equation}
        R_{{\mathrm e}, {\boldsymbol{s}} }   =  \mathfrak{r} \, g^2 k^2 \left|\int_0^{x_f} dx \;x^{-i{\omega} /\kappa} e^{-i s x} \right|^2\;,
    \label{eq:near-horizon-pex}
\end{equation}
where $s=C{\omega} + \nu/e $, and $C$ is given by Eq.~(\ref{eq:constant-C}). The integrand in Eq.~(\ref{eq:near-horizon-pex}) consists of two oscillatory factors $x^{-i{\omega} /\kappa}$ and $e^{-isx}$. The $e^{isx}$ factor is highly oscillating in the limit $\nu\gg {\omega} $, thus making the integral average out to essentially zero, away from the horizon---this is because the $x^{-i{\omega} /\kappa}$ factor barely changes over multiple oscillations of the function $e^{isx}$. However, near the horizon, the $x^{-i{\omega} /\kappa}$ factor exhibits a remarkable scale invariance that is a signature of the CQM modes. The important parameter that governs this dominant conformal property is ${\omega} /\kappa = 2\Theta$, which includes one contribution of 
$\Theta$ from the field modes piling up near the horizon, and the other one arising from the relative motion of the atom with respect to the fields. Now, because of the scale invariance of this factor 
$x^{-2 i \Theta}= x^{-i \omega/\kappa}$,
 as one zooms in towards the horizon, its behavior remains unaltered. As a result, the oscillations of $e^{isx}$ are no longer effective in rendering a vanishing contribution to the integral, and this generates the leading nonzero value from the near-horizon region. 

Due to the cancellation of the contributions from the integral away from the horizon, one can also increase the upper limit from $x_f$ to infinity without adding much error. The conformal behavior of the $x^{-i{\omega} /\kappa}$ factor ensures that the integral is nonzero, and captures the correct final results. Consequently, the emission rate is evaluated as
\begin{equation}
    R_{{\mathrm e}, {\boldsymbol{s}} }  =  \mathfrak{r} \,
    g^2 k^2 \left|\int_0^{x_f} \; dx\;  x^{-i{\omega} /\kappa} e^{-i{s} x} \right|^2  
    \longrightarrow  \mathfrak{r}  \, g^2 k^2  \left|\int_0^{\infty} \; dx\;  x^{-i{\omega} /\kappa} e^{-i{s} x} \right|^2 = \frac{2\pi  \mathfrak{r} \,  g^2 {\omega} }{\kappa\,\nu^2}\;\frac{1}{e^{2\pi{\omega} /\kappa}-1}\;,
    \label{eq:R_ex_steps} 
\end{equation}
where we have used the condition $\nu\gg {\omega} $. 
In this final result~(\ref{eq:R_ex_steps}), the constants $k$ and $C$
have disappeared, thus leading to an outcome that is independent of the initial conditions.

It should be noted that, if we had used a purely ingoing instead of a purely outgoing component, the emission rate would have been zero. This is because, for ingoing waves, the logarithmic contributions from the coordinate time $t$ and the field modes would have canceled each other out.
In other words, even though the near-horizon expressions of the modes involve different CQM combinations, only the purely outgoing components 
$  \Phi^{+ (CQM)}_{{\boldsymbol{s}} } 
     \stackrel{(\mathcal H)}{\propto}
     x^{ i\Theta} Y_{lm}(\Omega) e^{-i\omega t} 
     $
     [see Eq.~(\ref{eq:Schwarzschild_CQM_modes})] 
     survive for the transition processes considered in this paper. 
An important consequence of this property is that acceleration radiation with a Planckian distribution from a freely falling atom will exist for the Boulware state $\left| B \right\rangle$, as a result of the nonzero conformal integral in Eq.~(\ref{eq:R_ex_steps}).

 The emergence of the Planck factor in Eq.~(\ref{eq:R_ex_steps}) is an indication of the thermal behavior of the emission rate. This thermality will be further explored in Sec.~\ref{sec:density-matrix_steady-state} with the field density matrix. We can further confirm this behavior by finding the ratio of the emission and absorption rates. The absorption rate can easily be obtained from the emission rate expression by using the substitution ${\omega}  \rightarrow -{\omega} $, as can be seen from Eq.~(\ref{eq:P_CQM}). This gives
\begin{equation}
    R_{{\mathrm a}, {\boldsymbol{s}} } 
     = \frac{2\pi r g^2 {\omega} }{\kappa\,\nu^2}\;\frac{1}{1-e^{-2\pi{\omega} /\kappa}}\;,
\end{equation}
so that the ratio is 
\begin{equation}
    \frac{R_{{\mathrm e}, {\boldsymbol{s}} } }{R_{{\mathrm a}, {\boldsymbol{s}} } } 
    = e^{-2\pi{\omega} /\kappa}
    \,.
     \label{eq:ratio_em_ab}
\end{equation}
This ratio can be interpreted as corresponding to a thermal state with an effective temperature 
$T= \beta^{-1}$
 defined from the detailed-balance Boltzmann factor
\begin{equation}
    \frac{R_{{\mathrm e}, {\boldsymbol{s}} } }{R_{{\mathrm a}, {\boldsymbol{s}} } } = e^{-\beta {\omega} }
    \, . \label{eq:ratio_em_ab_Boltzmann}
\end{equation}
For the remainder of the paper, we will use units with the Boltzmann constant $k_{B} =1$.
Comparison of the right-hand sides of
 Eqs.~(\ref{eq:ratio_em_ab}) and (\ref{eq:ratio_em_ab_Boltzmann}) 
 gives the temperature of the thermal  state of the field as 
\begin{equation}
T= \beta^{-1}  = \frac{\kappa}{2\pi} \equiv \beta_{H}^{-1} = T_H
\; ,
\label{eq:temperature=Hawking}
\end{equation}
which is the Hawking temperature of the black hole. Moreover, this temperature coincides with the one predicted from the Planck distribution of Eq.~(\ref{eq:R_ex_steps}). 
A proof of the consistency of these statements using the steady-state condition for the density matrix is given in Sec.~\ref{sec:density-matrix_steady-state}, where we will achieve a complete characterization of the state as thermal.

Two critically important properties can be deduced from the derivation leading to 
 Eqs.~(\ref{eq:ratio_em_ab})--(\ref{eq:temperature=Hawking}).
 First, these equations
  show the existence of a {\it unique temperature $T$ defined uniformly for all modes by the Boltzmann factor\/} of Eq.~(\ref{eq:ratio_em_ab_Boltzmann}). If this unique-temperature condition were not satisfied, the ``temperature'' would be merely an effective phenomenological parameter introduced in an ad hoc manner for a particular mode. Instead, we have derived this uniqueness for the general class of black holes considered in this paper, with the implication that
$
T = T_{H}
$
is a candidate for a {\it genuine thermodynamic temperature associated with a thermal state\/}.
Second, the existence 
of this nontrivial uniqueness condition can be traced to the governing role played by the near-horizon region of the black hole through CQM, which yields the Boltzmann factor from the logarithmic singular nature of its modes. This happens in a way that it is encoded by the ratio of the probabilities $P_{{\mathrm e}, {\boldsymbol{s}} } $ and $P_{{\mathrm a}, {\boldsymbol{s}} } $, as displayed by Eq.~(\ref{eq:ratio_em_ab_Boltzmann}).
In conclusion, the anticipated {\it thermal state of the field and the Hawking temperature are completely determined by near-horizon CQM\/}. However, in order to accomplish an even more thorough thermal characterization of the state of the field, and to further understand the details of the radiation genesis, we will derive the steady-state field density matrix.

\section{Steady-state solution of the reduced field density matrix and complete thermal characterization}\label{sec:density-matrix_steady-state}
In Sec.~\ref{sec:master_equation}, we introduced the master equation~(\ref{eq:master_equation_final_multimode}) in its diagonal form, valid for random atomic injection times. We are particularly interested in the steady-state solution defined by the condition that
  the coarse-grained time derivative in Eq.~(\ref{eq:master_equation_final_multimode}) is equal to zero.
 The density matrix of this steady state will be denoted with superscript $(SS)$, namely,
 $  {{\rho}}^{\mathrm (SS)}_{\rm diag}(  \boldsymbol{ \left\{  \right. } n  \boldsymbol{\left. \right\}  } )$. From the results of Sec.~\ref{sec:conformal_steady_state}, especially Eqs.~(\ref{eq:ratio_em_ab}) and (\ref{eq:temperature=Hawking}), we can anticipate that this is a candidate for a thermal state. However, it is crucial to gain further insights into this behavior by a more thorough characterization of the thermal property via the master equation.
   
As a first step, it is straightforward to find the solution for the simpler single-mode equation~(\ref{eq:master_equation_final_single-mode}). In this case, the right-hand side gives a homogeneous three-term  linear recurrence relation that admits a power-law solution
${\rho}_{n,n}^{\mathrm (SS)} 
=  N \bigl(R_{{\mathrm e}, {\boldsymbol{s}} }/R_{{\mathrm a}, {\boldsymbol{s}} }\bigr)^n$, 
as can be easily verified.
The constant $N$ is determined from the normalization condition $\mathrm{Tr} \, [{\rho}] = 1$, which is a geometric series, $  \sum_{n=0}^{\infty}     {\rho}_{n,n}^{\mathrm (SS)}  = 
1 \implies N 
 \left[ 1-\left(R_{{\mathrm e}, {\boldsymbol{s}} }/R_{{\mathrm a}, {\boldsymbol{s}}} \right)  \right]^{-1} = 1$. 
 Thus,
\begin{equation}
\left. {\rho}_{n,n}^{\mathrm (SS)} \right|_{\mathrm{single-mode}}
 = 
  \left[ 1-\left( \frac{ R_{{\mathrm e}, {\boldsymbol{s}} } }{ R_{{\mathrm a}, {\boldsymbol{s}}} } \right)  \right]
  \,
 \left( \frac{R_{{\mathrm e}, {\boldsymbol{s}} } }{ R_{{\mathrm a}, {\boldsymbol{s}} }} \right)^n
    \; ,
    \label{eq:steady_state_single-mode}
    \end{equation}
given the expected inequality $R_{{\mathrm e}, {\boldsymbol{s}} } <R_{{\mathrm a}, {\boldsymbol{s}} }$. 
 In order to facilitate the transition to the multimode case, we can write this expression more precisely as 
${\rho}_{n_{j},n_{j}}^{\mathrm (SS)} =  N_{j} \bigl(R_{{\mathrm e}, j }/R_{{\mathrm a}, j} \bigr)^{n_{j}}$,  
for mode $n_{j}$, 
with $N_{j} = Z_{j}^{-1} 
= 1-\bigl(R_{{\mathrm e}, j }/R_{{\mathrm a}, j }\bigr)$.

As in Sec.~\ref{sec:master_equation}, the more relevant multimode master equation~(\ref{eq:master_equation_final_multimode}) needs to be used for a proper treatment of the quantum field thermodynamics. As pointed out, the modes are effectively independent if the atom-injection distribution is random. Thus, one can propose the ansatz that the complete steady-state multimode density matrix 
$ {\rho}^{\mathrm (SS)}$ will be factorized in terms of the tensor product of the density matrices
$ {\rho}^{\, j \, \mathrm (SS)} $ of the individual modes:
$
 {\rho}^{\mathrm (SS)}
 =
{\displaystyle 
\bigotimes_{j} 
} 
\, {\rho}^{\, j \, \mathrm (SS)} $.
 Therefore, in particular, each multimode diagonal element is the product of single-mode diagonal elements, such that
\begin{equation}
 {\rho}_{\rm diag}^{\mathrm (SS)}(  \boldsymbol{ \left\{  \right. } n  \boldsymbol{\left. \right\}  } )
 =
 \prod_{j} {\rho}_{n_{j},n_{j}}^{\mathrm (SS)} 
 \; ,
 \label{eq:SS-density-matrix_factorization}
\end{equation}
where ${\rho}_{n_{j},n_{j}}^{\mathrm (SS)} \equiv {\rho}_{n_{j},n_{j}}^{\mathrm \, j \, (SS)} $, i.e., for each mode $j$, the same functional form~(\ref{eq:steady_state_single-mode}) applies.
Then, for the steady state generated by
Eq.~(\ref{eq:master_equation_final_multimode}),
\begin{equation}
\begin{aligned}
0 =
    - 
     \sum_{j}
     \left(  \prod_{k \neq j} {\rho}_{n_{k},n_{k}}^{\mathrm (SS)}        
     \right)
     &
     \left\{
     R_{{\mathrm e},\, j}  \big[(n_j+1) \,
   {{\rho}}^{\mathrm (SS)}_{n_{j},n_{j}}
      - n_j \,
      {{\rho}}^{\mathrm (SS)}_{n_{j}-1 ,n_{j}-1 }
            \big] \right.
          \\
           &
            \left.
            +  
        R_{{\mathrm a},\, j} \big[ n_j \,
   {{\rho}}^{\mathrm (SS)}_{n_{j},n_{j}}
      - (n_j + 1)  \,
       {{\rho}}^{\mathrm (SS)}_{n_{j}+1 ,n_{j}+1 }
                           \big] \right\}
    \label{eq:master_equation_final_multimode_SS}
    \; ,
\end{aligned}
\end{equation}
where each term (labeled by $j$) 
has the form of the right-hand side of the single-mode equation~(\ref{eq:master_equation_final_single-mode}). 
Accordingly, the right-hand side of
Eq.~(\ref{eq:master_equation_final_multimode_SS}) 
vanishes term by term
because 
${{\rho}}^{\mathrm (SS)}_{n_{j},n_{j}} \equiv
    \left. {\rho}_{n,n}^{\mathrm (SS)} \right|_{\mathrm{single-mode}}
$ (with $n\equiv n_{j}$),
given by~(\ref{eq:steady_state_single-mode}),
 is the steady-state solution of the single-mode equation~(\ref{eq:master_equation_final_single-mode}).
 This sequential argument shows that the steady-state multimode reduced field density matrix is
 \begin{equation}
 {{\rho}}^{\mathrm (SS)}_{\rm diag}(  \boldsymbol{ \left\{  \right. } n  \boldsymbol{\left. \right\}  } )
=   \prod_{j} \left[ N_{j} \left( \frac{R_{{\mathrm e}, j } }{ R_{{\mathrm a}, j} } \right)^{n_{j}} \right]
= N \, \prod_{j}  \left(  \frac{R_{{\mathrm e}, j } }{ R_{{\mathrm a}, j} } \right)^{n_{j}} 
\label{eq:steady_state_multimode}
\; ,
\end{equation}
where $N=  \prod_{j} N_{j}  = \prod_{j} \left[ 1-\bigl(R_{{\mathrm e}, j }/R_{{\mathrm a}, j }\bigr) \right] $.
This automatically guarantees the trace normalization condition
$ \mathrm{Tr} \left[  {\rho}^{\mathrm (SS)} \right] = 1$, 
where the trace is computed with 
$\mathrm{Tr} \left[  \; \right]
=
\sum_{
 \boldsymbol{ \left\{  \right. } n  \boldsymbol{\left. \right\} }
 } \left[  \; \right]_{\rm diag}
 = \sum_{n_{ \boldsymbol{s}_{1}}} \sum_{n_{ \boldsymbol{s}_{2}}} \ldots
\left[  \; \right]_{\rm diag} $.

The closed-form solution~(\ref{eq:steady_state_multimode})
of the steady-state multimode density matrix can be interpreted as representing a thermal state with an effective temperature $T =\beta^{-1}$ defined from the detailed-balance Boltzmann factor
 \begin{equation}
 e^{-\beta {\omega} _{j}} = \frac{R_{{\mathrm e},j}}{R_{{\mathrm a},j}} 
 = \frac{P_{{\mathrm e},j}}{P_{{\mathrm a},j}}
 \; ,
 \label{eq:Boltzmann-factor}
 \end{equation}
  where ${\omega}  \equiv {\omega} _{j}$ is the positive frequency associated with the given field mode labeled by $j$. Of course, Eq.~(\ref{eq:Boltzmann-factor}) is the same as Eq.~(\ref{eq:ratio_em_ab_Boltzmann}), but we are now verifying that the Boltzmann detailed-balance condition is a restatement of the more general steady-state condition of the field master equation. Thus, Eqs.~(\ref{eq:steady_state_multimode}) and (\ref{eq:Boltzmann-factor}) generate the manifestly thermal form of the steady-state reduced field density matrix 
\begin{equation}
   {{\rho}}^{\mathrm (SS)}_{\rm diag}(  \boldsymbol{ \left\{  \right. } n  \boldsymbol{\left. \right\}  } ) = 
   \prod_{j}  
   \left[ e^{-n_{j} \beta {\omega} _{j}} (1-e^{- \beta {\omega} _{j}}) \right]
   = \frac{1}{Z}
   \prod_{j}  
    e^{-n_{j} \beta {\omega} _{j}}
    \; ,
    \label{eq:steady_state_multimode_thermal}
\end{equation}
where $Z= N^{-1}= 
\prod_{j} Z_{j}
= \prod_{j} \left[ 1-\bigl(R_{{\mathrm e}, j }/R_{{\mathrm a}, j }\bigr) \right]^{-1} $.
In this factorization, the single-mode density matrix is 
\begin{equation}
\left. {\rho}_{n_{j},n_{j}}^{\mathrm (SS)} \right| =  \frac{1}{ Z_{j}} \, e^{-n_{j} \beta {\omega} _{j}}
\; ,
\label{eq:steady_state_single-mode_thermal}
\end{equation}
 with $Z_{j} = N_{j}^{-1}$.
These are the familiar expressions that completely characterize a thermal state, including all moments of the probability distribution; in particular, the average steady-state occupation numbers per mode are given by the familiar Planck distribution
\begin{equation}
 \left\langle n_{j} \right\rangle \! ^{\! \! ^{\mathrm (SS)}}
   \equiv \mathrm{Tr} \left[   n_{j}  {{\rho}}^{\mathrm (SS)}_{\rm diag}(  \boldsymbol{ \left\{  \right. } n  \boldsymbol{\left. \right\}  } )  \right]
  =
\sum_{
 \boldsymbol{ \left\{  \right. } n  \boldsymbol{\left. \right\} }
}
n_{j} \,   {{\rho}}^{\mathrm (SS)}_{\rm diag}(  \boldsymbol{ \left\{  \right. } n  \boldsymbol{\left. \right\}  } ) 
  =
  \sum_{n_{j}=0}^{\infty} n_{j} \, {\rho}_{n_{j},n_{j}}^{\mathrm (SS)} = \frac{1}{ e^{\beta {\omega} _{j}} -1}
    \; .
    \label{eq:average-occupation_thermal}
\end{equation}
In Eq.~(\ref{eq:average-occupation_thermal}),
the trace is computed as discussed after Eq.~(\ref{eq:steady_state_multimode}), 
and the result follows from the factorization~(\ref{eq:steady_state_multimode_thermal}) and normalization conditions for all the single-mode density matrices. The Planck distribution~(\ref{eq:average-occupation_thermal}) for the occupation number averages is in agreement with the earlier result for the radiation probability, Eq.~(\ref{eq:R_ex_steps}).

A crucial point in the derivation of the thermal characterization of the state of the quantum field, Eqs.~(\ref{eq:steady_state_multimode_thermal}) and (\ref{eq:average-occupation_thermal}), is again the requirement of a {\it unique temperature defined uniformly by the Boltzmann factor\/} of Eq.~(\ref{eq:Boltzmann-factor}). The existence of a unique temperature---which has a conformal character driven by CQM, as shown in Sec.~\ref{sec:conformal_steady_state}---and the density-matrix results of this section prove that the field is described by a thermal state near its steady-state configuration. 
The results hold under the diagonal property of the density matrix leading to the 
master equation~(\ref{eq:master_equation_final_multimode}, e.g., due to random injection times of the atomic cloud. This remarkable finding has additional thermodynamic consequences, as will be briefly discussed in the next section, along with closing remarks.

\section{Implications and outlook} \label{sec:conclusion}
Our detailed analysis of the field equations, geodesics, and density matrix 
proves the thermal nature of the steady state of the acceleration radiation field (HBAR)
due to time-randomly injected freely falling atoms in the Boulware state of the generalized Schwarzschild black hole. 
This confirms the validity of the results of Ref.~\cite{scully2018}, in a more general setting and including a detailed simultaneous accounting of all the field modes.  
Most importantly, the thermal property of the field density matrix is completely governed by the conformal near-horizon physics of the black hole. This is due to the fact that the field modes near the horizon are described by the CQM Hamiltonian; and the scale symmetry of these modes is responsible for the thermality of the HBAR field.
This conformal thermal nature involves both the Hawking temperature and additional thermodynamic consequences---remarkably, they both appear to behave in ways analogous to the thermodynamic properties of the black hole itself.

In principle, a complete thermodynamic framework can be established by computing the entropy 
directly from the density matrix in its quantum von Neumann form. 
As a back-of-the-envelope calculation to anticipate the result,
one can consider the single-mode density matrix~(\ref{eq:master_equation_final_single-mode})
as in Ref.~\cite{scully2018},
whence (by straightforward algebra), for a generalized Schwarzschild metric,
\begin{equation}
    \dot{S}_{\mathcal P} 
    =
    \beta_{H}
    \dot{\, \, \left\langle n_{ \boldsymbol{s} } \right\rangle }
\,   \omega
    = \beta_{H}
 \dot{E}_{\mathcal P}
    \; ,
    \label{eq:Sp_2}
    \end{equation}
where both
$ \dot{S}_{\mathcal P} $
and
$ \dot{E}_{\mathcal P}$
should be properly interpreted as sums over all the modes.
As a result, using the Hawking temperature~(\ref{eq:temperature=Hawking})
and its relationship with the black hole parameters (mass $M$ or related $r_+$),
one concludes that the HBAR entropy is given by
\begin{equation}
    \dot{S}_{\mathcal P} = \frac{1}{4} \bigl| \dot{A}_{\mathcal P} \bigr|
     \label{eq:HBAR_final}
     \; ,
\end{equation}
where $\bigl| \dot{A}_{\mathcal P} \bigr|$ is the 
change in the event horizon area 
associated with HBAR emission of photons (with the absolute value to account for the signs).
This result, which is structurally identical to the Bekenstein-Hawking entropy of the black hole itself,
 will be rigorously justified in the second paper of this series,
where a larger thermodynamic correspondence between the HBAR field and black hole thermodynamics will be fully developed~\cite{HBAR_part-II}.
For example, for the generalized Schwarzschild geometry, this shows that there is deep connection between the HBAR radiation field and the black hole,
with a correspondence
\begin{equation}
\bigl( 
{S}_{\mathcal P} , {E}_{\mathcal P} 
\bigr) 
\xlongleftrightarrow{\beta = \beta_{H}}
\bigl( 
S_{\mathrm{BH}} , M 
\bigr) 
\; \; \;
\; ,
\label{eq:HBAR-BH-correspondence}
\end{equation}
which is established at the common Hawking temperature
 $T_{H}=\beta_{H}^{-1}$.
Moreover, as will be discussed in~\cite{HBAR_part-II}, these analog relations can be further extended to rotating and charged black holes (Kerr-Newman geometry), with angular momentum and charge variables. 

In closing, this article has analyzed the essential features of the acceleration radiation for an atomic cloud falling into a Schwarzschild black hole,
and it paves the way for a more comprehensive thermodynamic framework.

\acknowledgments{}
H.E.C. acknowledges support by the University of San Francisco Faculty Development Fund. 
ThM.O.S.\ and A.A.\ acknowledge support by the
National Science Foundation (Grant No.\ PHY-2013771), the Air Force Office of Scientific Research (Award No.\ FA9550-20-1-0366 DEF), the Office of Naval Research (Award No.\ N00014-20-1-2184), the Robert A. Welch Foundation (Grant No.\ A-1261) and the King Abdulaziz City for Science and Technology (KACST).
H.E.C. acknowledges support
by the University of San Francisco Faculty Development Fund. 
This material is based upon work supported by the Air Force Office of Scientific Research under award FA9550-21-1-0017 (C.R.O. and A.C.).   

\begin{appendix}
\section{Derivation of the master equation for the field---Eqs.~(\ref{eq:master_equation_final_single-mode})
and (\ref{eq:master_equation_final_multimode})---and related general properties}
\label{app:master_equation_derivation}
In this appendix, we provide the details of the derivation of the field master equation. We will start by showing the basic algebra that leads from Eq.~(\ref{eq:masterequation1}) to its single-mode version~(\ref{eq:master_equation_final_single-mode}), and focusing on the diagonal terms of the reduced field density matrix. This will be followed by a discussion of the single-mode off-diagonal terms and the general structural form of the multimode version.
In all cases, we will consider atoms that are injected in the ground state: $ \rs{\mathcal A} = \ket{b}\bra{b}$.
\subsection{Generic form of the master equation in terms of field operators}
\begin{figure}[h]
    \centering
    \includegraphics[width=0.4\linewidth]{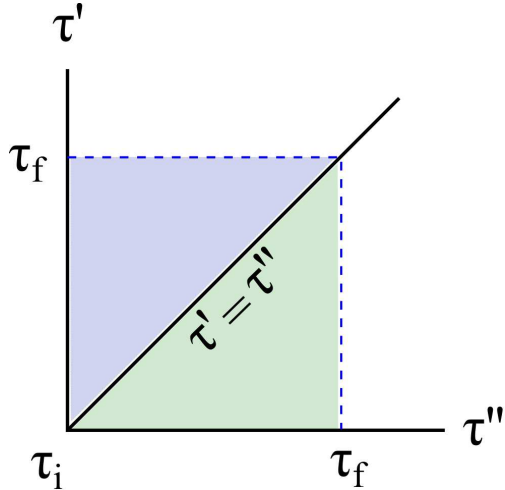}
    \caption{Regions covered by the two integration limits in Eq.~(\ref{eq:int-trace-2}).}    \label{fig:limit_integration}
\end{figure}

As shown in the main text, the right-hand side of the field master equation~(\ref{eq:masterequation1}) involves a double integral of the partial trace of double commutators~(\ref{eq:int-trace-commutator}). When the replacements~(\ref{eq:commutator-VVrho}) and (\ref{eq:atomic-trace}) are made in Eq.~(\ref{eq:int-trace-commutator}), with the upper limit $\tau_f =\tau$,
 \begin{equation}
 \begin{aligned}
  \! \! \!   \int_{\tau_i}^{\tau_f} \int_{\tau_i}^{\tau'}
   \! \! d\tau'd\tau''\,\,\mathrm{Tr_{\mathcal A}} 
   &
   \left[ V', [V'', \rs{{\mathcal P}{\mathcal A}}  ] \right] 
    = 
g^2  \int\displaylimits_{\tau'>\tau''} 
d^2 \tau \,  \left(  e^{- i  \nu \tau'} e^{ i \nu \tau''}
\Phi' \Phi'' \, {\rho}
 +  e^{ i  \nu  \tau'} e^{ -i \nu \tau''}
{\rho} \, \Phi''  \Phi'
\right)
\\
& -
g^2
 \int\displaylimits_{\tau'>\tau''} 
d^2 \tau  \,    
\left( e^{ i  \nu \tau'} e^{ - i \nu \tau''} \Phi' \, {\rho} \, \Phi'' 
 +  e^{- i \nu  \tau'} e^{  i \nu \tau''} \Phi'' \,  {\rho} \, \Phi'  \right)
    \label{eq:int-trace-2}
    \; ,
\end{aligned}
\end{equation}
where
$d^2 \tau = d \tau' d \tau''$, and the region of integration is the domain 
$I  = \left\{ \tau_i \leq \tau'' \leq \tau' ; \tau_i \leq \tau' \leq \tau_f \right\} $, which we represent loosely as 
$I  = \left\{ \tau' > \tau'' \right\} $; referring to Fig.~\ref{fig:limit_integration}, this region
is the blue-highlighted upper triangle. In each parenthesis of Eq.~(\ref{eq:int-trace-2}), the second term corresponds to the adjoint in Eq.~(\ref{eq:commutator-VVrho}). (This operation also includes the complex conjugates of the exponentials because the latter arise from the adjoint of the atomic operators.)
A more compact and convenient version of Eq.~(\ref{eq:int-trace-2}) can be obtained by renaming the variables  in each of the second terms within the parentheses; this can be performed via the exchange of symbols 
$\tau' \leftrightarrow \tau''$ that formally converts region I into region 
$II  = \left\{ \tau_i \leq \tau' \leq \tau'' ; \tau_i \leq \tau'' \leq \tau_f \right\} $ 
(which we represent loosely as $II  = \left\{ \tau' < \tau'' \right\} $). 
In  Fig.~\ref{fig:limit_integration}, region II is the green-highlighted lower triangle. 
Thus, Eq.~(\ref{eq:int-trace-2}) becomes
 \begin{equation}
 \begin{aligned}
  \! \! \!    \int_{\tau_i}^{\tau_f} \int_{\tau_i}^{\tau'} 
   \! \! d\tau'd\tau''
   \,\,\mathrm{Tr_{\mathcal A}} 
   & \left[ V', [V'', \rs{{\mathcal P}{\mathcal A}}  ] \right] 
    = 
  g^2  \left( \;
 \int\displaylimits_{\, I} 
d^2 \tau \,    e^{- i   \nu \tau'} e^{ i \nu \tau''}
\Phi' \Phi'' \, {\rho}
 +  
 \int\displaylimits_{II}  
 e^{ -i  \nu \tau'} e^{ i \nu \tau''}
{\rho} \, \Phi'  \Phi''
\right)
\\
 &   - g^2
 \left( \;
 \int\displaylimits_{\, I} 
d^2 \tau  \,    
e^{ i  \nu \tau'} e^{ - i \nu \tau''} \Phi' \, {\rho} \, \Phi''  + 
\int\displaylimits_{ II } 
e^{ i  \nu \tau'} e^{-  i \nu \tau''} \Phi' \, {\rho} \, \Phi'' 
\right)
    \; ,
    \label{eq:int-trace-3}
\end{aligned}
\end{equation}
which can be recast into the form of Eq.~(\ref{eq:masterequation2}) in the main text.
The main advantage gained by this coordinate relabeling is that it effectively leads to the coverage of the entire square region $I+II$: in effect, for the second group of integrals, the integrand is symmetric and it is to be integrated uniformly over the square region; but, for the first group of integrals, this simplification only holds for the diagonal elements, as shown below.

\subsection{Single-mode master equation}
We are now ready to turn the master equation into the powerful form used in quantum optics~\cite{scullybook}. We will write the coarse-grained rate of change of the diagonal elements of the reduced field density matrix
$  {d \rs{\mathcal P}}/{d t} $, using Eq.~(\ref{eq:masterequation2}).
 This can be done by considering the three generic matrix elements in Eq.~(\ref{eq:masterequation2}):
 $\bra{n}  \Phi' \, \Phi'' {\rho}  \, \ket{n}$,
 $\bra{n}  {\rho} \, \Phi' \Phi'' \ket{n}$,
 and
 $\bra{n}  \Phi' \, {\rho} \, \Phi''   \ket{n}$.

For the single-mode version of the reduced field density matrix, the field operator is 
$\Phi = a_{\boldsymbol{s}} \phi_{\boldsymbol{s}} ( {\mathbf r}, t) + a_{\boldsymbol{s}}^{\dagger} \phi_{\boldsymbol{s}}^{*} ( {\mathbf r}, t)$, 
which can be succinctly written as $\Phi = a \phi + a^{\dagger} \phi^{*} $. 
Therefore, the three generic matrix elements are
 \begin{equation}
 \begin{aligned}
 \bra{n}  \Phi' \Phi'' \, {\rho}  \ket{n}
 & = 
 n \, {\rho}_{n,n} \phi'^{*} \phi''
 +  (n+1) {\rho}_{n,n} \phi' \phi''^{*} 
 \\
& + \sqrt{(n+1)(n+2)} \, {\rho}_{n+2,n}\, \phi' \phi''
 + \sqrt{n(n-1)} \, {\rho}_{n-2,n} \, \phi'^{*} \phi''^{*}
   \\
 \bra{n}  {\rho} \, \Phi' \Phi'' \ket{n}
   & = 
 n \,  {\rho}_{n,n} \, \phi'^{*} \phi''
 +  (n+1) \, {\rho}_{n,n} \,\phi' \phi''^{*} 
 \\
 & + \sqrt{(n+1)(n+2)} \, {\rho}_{n,n+2}\,  \phi'^{*} \phi''^{*}
 + \sqrt{n(n-1)} \, {\rho}_{n, n-2} \, \phi' \phi''
  \\
 \bra{n}  \Phi' \, {\rho} \, \Phi''   \ket{n}
  & = 
 n \, {\rho}_{n-1,n-1} \phi'^{*} \phi''
 +  (n+1) \, {\rho}_{n+1,n+1} \phi' \phi''^{*} 
 \\
 &
 + \sqrt{n(n+1)} \, {\rho}_{n+1,n-1}  \,\phi' \phi''
 + \sqrt{n(n+1)} \, {\rho}_{n-1, n+1} \, \phi'^{*} \phi''^{*}
 \; .
\end{aligned}
\label{eq:ME-generic-matrix-elements}
\end{equation}
In Eq.~(\ref{eq:ME-generic-matrix-elements}), the following more primary results were used (as a direct consequence of the properties of creation and annihilation operators):
 \begin{equation}
 \begin{aligned}
  & \bra{n}  a_{\boldsymbol{s}}^\dagger a_{\boldsymbol{s}} \, {\rho} \, \ket{n} = n \, {\rho}_{n,n} = \bra{n} \, {\rho} \, a_{\boldsymbol{s}}^\dagger a_{\boldsymbol{s}} \ket{n}
   &
     \bra{n} a_{\boldsymbol{s}} a_{\boldsymbol{s}}^\dagger \, {\rho} \, \ket{n} = (n+1) \, {\rho}_{n,n} = \bra{n} \, {\rho} \, a_{\boldsymbol{s}} a_{\boldsymbol{s}}^\dagger \ket{n}
   \\
   & \bra{n} a_{\boldsymbol{s}} a_{\boldsymbol{s}} \, {\rho} \, \ket{n} = \sqrt{(n+1)(n+2)} \, {\rho}_{n+2,n} 
    & 
     \bra{n}  a^\dagger_{\boldsymbol{s}} a^\dagger_{\boldsymbol{s}} \, {\rho} \, \ket{n} =\sqrt{n(n-1)} \,  {\rho}_{n-2,n} 
     \\
    &   \bra{n} \, {\rho} \, a^\dagger_{\boldsymbol{s}} a^\dagger_{\boldsymbol{s}} \ket{n} =\sqrt{(n+1)(n+2)} \,  {\rho}_{n,n+2} 
   &
      \bra{n}  {\rho} \, a_{\boldsymbol{s}} a_{\boldsymbol{s}}  \ket{n} =\sqrt{n(n-1)} \,  {\rho}_{n,n-2} 
   \\
      & \bra{n}  a^\dagger_{\boldsymbol{s}} \, {\rho} \, a_{\boldsymbol{s}}  \ket{n} = n \, {\rho}_{n-1,n-1} 
       &
       \bra{n}  a_{\boldsymbol{s}} {\rho} \, a^\dagger_{\boldsymbol{s}}  \ket{n} = (n+1) \, {\rho}_{n+1,n+1}   
       \\
     &  \bra{n}  a_{\boldsymbol{s}} \, {\rho} \, a_{\boldsymbol{s}}  \ket{n} = \sqrt{n(n+1)} \, {\rho}_{n+1,n-1} 
       &
       \bra{n}  a^\dagger_{\boldsymbol{s}} \, {\rho} \, a^\dagger_{\boldsymbol{s}}  \ket{n} = \sqrt{n(n+1)} \, {\rho}_{n-1,n+1}   
        \; .
    \end{aligned}
 \label{eq:ME-primary-matrix-elements}   
\end{equation}
It should be noted that the expanded products in terms of creation and annihilation operators lead to twelve terms from Eq.~(\ref{eq:ME-generic-matrix-elements}), of which ten are independent with additional Hermitian symmetries. The basic structure of Eq.~(\ref{eq:masterequation2}) involves a set of terms with two factors each: the matrix elements~(\ref{eq:ME-primary-matrix-elements}), and the integrals 
\begin{equation}
\mathcal{J}^{(\mathrm{sgn} (\pi) )}_{(\epsilon',\epsilon'')} 
= \int\displaylimits_{I}  d^{2} \tau \,  e^{- i \, \mathrm{sgn} ( \pi) \, \nu \tau'} e^{ i \, \mathrm{sgn} (\pi) \, \nu  \tau''} \phi'_{(\epsilon')} \, \phi''_{(\epsilon'')}
\; ,
\label{eq:integral-structural}
\end{equation}
 where $\phi_{(\epsilon)}$, with $\epsilon = \pm$, are functions selected from $\phi$ and $\phi^{*}$ according to the convention $\phi_{(\mp)} = \phi, \phi^{*}$ (in that order). The correct assignments of functions and signs can be read off from Eqs.~(\ref{eq:ME-generic-matrix-elements}) and (\ref{eq:masterequation2}). 
The property 
$\left[ \mathcal{J}^{(\mathrm{sgn} (\pi) )}_{(\epsilon',\epsilon'')}  \right]^{*} =
 \mathcal{J}^{(-\mathrm{sgn} (\pi) )}_{(-\epsilon',-\epsilon'')} $
 can be verified by direct inspection.

If we assume that we can ignore the off-diagonal terms due to random injection averaging, we can then focus on the four terms that correspond to the first lines of 
 $\bra{n}  \Phi' \Phi'' \, {\rho}  \, \ket{n}$,
 $\bra{n}  {\rho} \, \Phi' \Phi'' \ket{n}$,
 and
 $\bra{n}  \Phi' \, {\rho} \, \Phi''   \ket{n}$ 
 in Eq.~(\ref{eq:ME-generic-matrix-elements}). The justification of the selection of diagonal elements due to random injection times will be analyzed in Appendix~\ref{app:injection-averages_diagonality}. 
 With this selection, the first and second terms on the right-hand side of
Eq.~(\ref{eq:int-trace-3}) become
  \begin{equation}
  \begin{aligned}
    &
    \bigl\langle n \bigr|
    \int_{\tau_i}^{\tau_f} \!  \int_{\tau_i}^{\tau'}
    \! d\tau'd\tau''  \!
    \left\{
    \mathrm{Tr_{\mathcal A}}[V'V''\rs{{\mathcal P}{\mathcal A}}] + \mathrm{h.c.}
    \right\} 
      \bigl| n \bigr\rangle
      =
      g^2 \! \left(
 \int\displaylimits_{\, I} 
d^2 \tau    \,  e^{- i   \nu \tau'} e^{ i \nu \tau''}
\bigl\langle n \bigr|
 \Phi' \Phi'' \, {\rho}
   \bigl| n \bigr\rangle
 + \mathrm{h.c.} 
\right)
    \\ & =   g^2  
     \int\displaylimits_{ I +II} \! 
     d^2 \tau \,    e^{- i   \nu \tau'} e^{ i \nu \tau''}
\bigl[ \phi'  \phi''^{*}  (n+1) \,  {\rho}_{n,n}
+
 \phi'^{*}  \phi'' n \, {\rho}_{n,n}
\bigr]
     \\
     & =   g^2  \! \left[
     \int_{\tau_i}^{\tau_f} \! d\tau' e^{-i\nu\tau'}  \phi' \!
    \int_{\tau_i}^{\tau_f} \! d\tau'' e^{i\nu\tau''} \phi''^{*} \right] \! (n+1) \, {\rho}_{n,n} + 
     g^2  \! \left[
      \int_{\tau_i}^{\tau_f} \! d\tau' e^{-i\nu\tau'}  \phi'^{*} \!
     \int_{\tau_i}^{\tau_f} \! d\tau'' e^{i\nu\tau''}  \phi''  \right] \! n \, {\rho}_{n,n} 
     \\ 
     & = 
     g^2 |I_{{\mathrm e},{\boldsymbol{s}}}|^2 (n+1) \, {\rho}_{n,n} + g^2 
     |I_{{\mathrm a},{\boldsymbol{s}}}|^2\, n \, {\rho}_{n,n}   =
    P_{{\mathrm e},{\boldsymbol{s}}} \, (n+1) \, {\rho}_{n,n} + P_{{\mathrm a},{\boldsymbol{s}}} \, n \, {\rho}_{n,n}  
    \; .
    \label{eq:first_term_final}
\end{aligned}
\end{equation}
and
\begin{equation}
  \begin{aligned}
    &  \! \! \! 
   - \bigl\langle n \bigr|
    \int_{\tau_i}^{\tau_f} \! \! \int_{\tau_i}^{\tau'}
    \! d\tau'd\tau''  \!
    \left\{
    \mathrm{Tr_{\mathcal A}}[V'\rs{{\mathcal P}{\mathcal A}}V''] + \mathrm{h.c.}
    \right\} 
      \bigl| n \bigr\rangle
      =
- g^2 \!
 \int\displaylimits_{\, I+II} 
d^2 \tau  \,    
e^{ i   \nu \tau'} e^{ - i \nu \tau''}  
\bigl\langle n \bigr| \Phi' \, {\rho} \, \Phi''  
 \bigl| n \bigr\rangle
  \\
  &   \! \! \!     
  = - g^2  
  \int\displaylimits_{ I +II} \! \!
d^2 \tau \,    
 e^{-  i   \nu \tau'} e^{  i \nu \tau''}  
\bigl[
 \phi'^{*} \phi'' n \, {\rho}_{n-1,n-1}
 +  \phi' \phi''^{*}  
  (n+1) \, {\rho}_{n+1,n+1}
 \bigr]
     \\
    &   
     \! \!       \! \! 
     = - g^2 \! \! 
    \left[ \!
     \int_{\tau_i}^{\tau_f} \! \! d\tau' e^{-i\nu\tau'}  \phi'^{*} 
     \! \! \!
    \int_{\tau_i}^{\tau_f} \! \! d\tau'' e^{i\nu\tau''}  \!
     \phi''  \right] \! \!
     (n+1)  {\rho}_{n+1,n+1}
     \!  - g^2 \! \!
      \left[ \!
      \int_{\tau_i}^{\tau_f} \! \! d\tau' e^{-i\nu\tau'}  \phi' 
       \! \! \!
     \int_{\tau_i}^{\tau_f} \! \! d\tau'' e^{i\nu\tau''} \phi''^{*} 
      \right] \!
       n  {\rho}_{n-1,n-1} 
     \\ &  \! \! \! 
    = -  g^2 |I_{{\mathrm a},{\boldsymbol{s}}}|^2 \, (n+1) \, {\rho}_{n+1,n+1} 
    -  g^2 |I_{{\mathrm e},{\boldsymbol{s}}}|^2 \, n \, {\rho}_{n-1,n-1}
    = -  P_{{\mathrm a},{\boldsymbol{s}}} \, (n+1) {\rho}_{n+1,n+1} 
    -  P_{{\mathrm e},{\boldsymbol{s}}} \, n \, {\rho}_{n-1,n-1}
    \,.
\end{aligned}
   \label{eq:second_term_final}
\end{equation}
 In the transition from the second to the third lines in Eqs.~(\ref{eq:first_term_final}) and (\ref{eq:second_term_final}), the double integrals, covering the whole square region $I+II$, admit the factorization into the product of independent integrals with respect to $\tau'$ and $\tau''$. The double integrals are probabilities factored into the product of probability amplitudes, as can be seen by direct comparison with Eqs.~(\ref{eq:P_ex_expression})--(\ref{eq:P_ab_expression})  and (\ref{eq:P_ex_explicit})--(\ref{eq:P_ab_explicit}). Moreover, these integrals are particular cases of the general pattern of Eq.~(\ref{eq:integral-structural}); specifically,
\begin{equation} 
\begin{aligned}
  P_{{\mathrm e},{\boldsymbol{s}}} 
  & = g^2 \left[ 
  \mathcal{J}^{( + )}_{(-,+)}  + \mathrm{c.c.} \right]
  = g^2 \left[ 
   \mathcal{J}^{( + )}_{(-,+)}  +   \mathcal{J}^{( - )}_{(+,-)} \right]
  \\
    P_{{\mathrm a}, {\boldsymbol{s}}}
    & = g^2 \left[ 
      \mathcal{J}^{( + )}_{(+,-)}  + \mathrm{c.c.} \right]
    = g^2 \left[ 
       \mathcal{J}^{( + )}_{(+,-)}  +   \mathcal{J}^{( - )}_{(-,+)} \right]
    \; ,
    \end{aligned}
    \label{eq:Pe-Pa_from_J-integrals}
    \end{equation}
where $\mathrm{c.c.}$ is the complex conjugate.

As a final step, combining all four terms from Eqs.~(\ref{eq:first_term_final}) and (\ref{eq:second_term_final})
into Eq.~(\ref{eq:masterequation2}),
we get the single-mode master equation for the field density matrix in terms of diagonal elements only, as given by Eq.~(\ref{eq:master_equation_final_single-mode}) in the main text. Furthermore, one can also compute the additional terms in Eq.~(\ref{eq:int-trace-2}) that give off-diagonal density matrix elements, by including the corresponding matrix elements from Eq.~(\ref{eq:ME-generic-matrix-elements}). 
Then, the more general single-mode master equation for the reduced field density matrix reads
\begin{equation}
\begin{aligned}
 \dot{\rho}_{n,n} = &
     - R_{{\mathrm e}, {\boldsymbol{s}} } \big[(n+1) \, {\rho}_{n,n} - n \, {\rho}_{n-1,n-1}\big]
      -  R_{{\mathrm a}, {\boldsymbol{s}} }\big[n \, {\rho}_{n,n}-(n+1) \, {\rho}_{n+1,n+1} \big] 
    \\
    &
   -
   \left[
    S_{+, {\boldsymbol{s}} } 
     \sqrt{(n+1)(n+2)} \, {\rho}_{n+2,n} 
    + S_{-, {\boldsymbol{s}} }  \sqrt{n(n-1)} \,  {\rho}_{n,n-2} 
    \right.
    \\
    &
    \left.
    - \left( S_{+, {\boldsymbol{s}} }  + S_{-, {\boldsymbol{s}} } \right)  \sqrt{n(n+1)} \, {\rho}_{n+1,n-1}       
    +  \mathrm{h.c.} 
\right]         
        \label{eq:master_equation_single-mode-squeezing}
    \; ,
\end{aligned}
\end{equation}
where 
\begin{equation}
\begin{aligned}
S_{\pm, {\boldsymbol{s}} } & = \mathfrak{r} \, g^2 J_{\pm, {\boldsymbol{s}} } 
\\
J_{\pm, {\boldsymbol{s}} } 
& =
 \int_{\tau_i}^{\tau_f} d\tau'\, 
 \int_{\tau_i}^{\tau'} d\tau'' 
 \, e^{\mp i\nu\tau'}  e^{\pm i\nu\tau''}
 \phi_{\boldsymbol{s}} (\mathbf{r} (\tau'),t(\tau')) \, \phi_{\boldsymbol{s}} (\mathbf{r} (\tau''),t(\tau''))
 \; .
 \label{eq:squeezing-coefficients}
 \end{aligned}
 \end{equation}
These integrals are also particular cases of the general pattern of Eq.~(\ref{eq:integral-structural}), with
  $J_{\pm, {\boldsymbol{s}} } 
  =
  \mathcal{J}^{( \pm )}_{(-,-)} 
  $.
The coefficients $S_{+, {\boldsymbol{s}} } $ and $S_{-, {\boldsymbol{s}} } $ are squeezing factors
that play a similar role as those used in laser theory~\cite{scully2003,scullybook}.
 As shown in the main text, Sec.~\ref{sec:density-matrix_steady-state}, the master equation~(\ref{eq:master_equation_final_single-mode}) leads to a thermal density matrix; by contrast, the more general master equation~(\ref{eq:master_equation_single-mode-squeezing}), when $S_{+, {\boldsymbol{s}} }, S_{-, {\boldsymbol{s}} } \neq 0$, leads to a non-thermal density matrix that generates squeezed radiation.

\subsection{Multimode master equation}
For the multimode version of the reduced field density matrix, the field operator is 
$\Phi = \sum_{{\boldsymbol{s}}} 
\left[ a_{\boldsymbol{s}} \phi_{\boldsymbol{s}} ( {\mathbf r}, t) 
+ a_{\boldsymbol{s}}^{\dagger} \phi_{\boldsymbol{s}}^{*} ( {\mathbf r}, t)
\right]$, 
which can be succinctly written as
$\Phi = \sum_{j} \left[ a_{j} \phi_{j} + a_{j}^{\dagger} \phi_{j}^{*} \right]$. 
Thus, the structural form of the master-equation terms giving the rate of change is similar to Eqs.~(\ref{eq:master_equation_final_single-mode}) and (\ref{eq:master_equation_single-mode-squeezing}),
with an additional summation over modes. This additional ingredient involves double sums $\sum_{j,k} $ 
because the right-hand side of Eq.~(\ref{eq:ME-generic-matrix-elements}) is quadratic in $\Phi$. These sums include mode-diagonal terms (contributions $j=k$) that are identical to those in Eq.~(\ref{eq:master_equation_single-mode-squeezing}); moreover, there are extra off-diagonal terms (contributions $j\neq k$) that involve creation and annihilation operators of different modes, as we will show next.

The three generic matrix elements have the expansions
 \begin{equation}
 \begin{aligned}
   \bra{n}  \Phi' \Phi''  \, {\rho}  \ket{n}
 & = 
 \sum_{j} 
 \biggl[
 n_{j} \, {\rho}_{n;n} \, \phi_{j} '^{*} \phi_{j} ''
 +  (n_{j}+1) \, {\rho}_{n;n}  \, \phi_{j} ' \phi_{j} ''^{*} 
 \\
&
+ \sqrt{(n_{j}+1)(n_{j}+2)} \, {\rho}_{n_{j}+2;n} \, \phi_{j} ' \phi_{j} ''
 + \sqrt{n_{j} (n_{j}-1)} \, {\rho}_{n_{j}-2;n} \, \phi_{j}'^{*} \phi_{j}''^{*}
 \biggr]
 \\
 &
 +  \sum_{j \neq k} 
 \biggl[
  \sqrt{ n_{j} (n_{k}+1 ) }  \, {\rho}_{n_{j}-1,n_{k}+1; n } \, \phi_{j} '^{*} \phi_{k} ''
 +   \sqrt{ (n_{j}+ 1 ) n_{k} }  \, {\rho}_{n_{j}+1,n_{k}-1; n }  \, \phi_{j} ' \phi_{k} ''^{*} 
 \\
& 
+ \sqrt{(n_{j}+1)(n_{k}+1)} \, {\rho}_{n_{j}+1,n_{k}+1;n} \, \phi_{j} ' \phi_{k} ''
 + \sqrt{n_{j} n_{k}} \, {\rho}_{n_{j}-1,n_{k}-1;n} \, \phi_{j}'^{*} \phi_{k}''^{*}
  \biggr]
  \end{aligned}
   \nonumber
   \end{equation}
  \begin{equation}
  \begin{aligned}
   \bra{n}  {\rho} \, \Phi' \Phi'' \ket{n}
   & = 
    \sum_{j} \left[
 n_{j} {\rho}_{n;n} \,  \phi_{j}'^{*} \phi_{j}''
 +  (n_{j}+1) {\rho}_{n;n} \, \phi_{j}' \phi_{j}''^{*} 
 \right.
  \\
 & \left.
  + \sqrt{(n_{j}+1)(n_{j}+2)} \, {\rho}_{n;n_{j}+2} \, \phi_{j}'^{*} \phi_{j}''^{*}
 +  \sqrt{n_{j} (n_{j}-1)} \, {\rho}_{n; n_{j}-2} \, \phi_{j}' \phi_{j}''
 \right]
  \\
  &
 +  \sum_{j \neq k} 
 \biggl[
  \sqrt{ (n_{j} + 1)  n_{k} }  \, {\rho}_{n; n_{j}+1,n_{k}-1 } \, \phi_{j} '^{*} \phi_{k} ''
 +   \sqrt{ n_{j} (n_{k} + 1 ) }  \, {\rho}_{n; n_{j}-1,n_{k}+1 }  \, \phi_{j} ' \phi_{k} ''^{*} 
 \\
& 
+ \sqrt{(n_{j}+1)(n_{k}+1)} \, {\rho}_{n; n_{j}+1,n_{k}+1 } \, \phi_{j}'^{*} \phi_{k}''^{*}
 + \sqrt{n_{j} n_{k}} \, {\rho}_{n; n_{j}-1,n_{k}-1 } \, \phi_{j} ' \phi_{k} ''
  \biggr]
    \end{aligned}
 \label{eq:ME-generic-matrix-elements_multimode}
  \end{equation}
  \begin{equation}
  \begin{aligned}
\bra{n}  \Phi' \, {\rho} \, \Phi''   \ket{n}
  & = 
  \sum_{j , k} 
  \biggl[
  \sqrt{ n_{j} n_{k} }  \, {\rho}_{n_{j}-1; n_{k}-1 } \, \phi_{j} '^{*} \phi_{k} ''
 +   \sqrt{ (n_{j}+ 1 ) (n_{k} +1)}  \, {\rho}_{n_{j}+1;n_{k}+1 }  \, \phi_{j} ' \phi_{k} ''^{*} 
 \\
& 
+ \sqrt{(n_{j}+1)n_{k} } \, {\rho}_{n_{j}+1;n_{k}- 1 } \, \phi_{j} ' \phi_{k} ''
 + \sqrt{n_{j} (n_{k} + 1) } \, {\rho}_{n_{j}-1; n_{k}+1 } \, \phi_{j}'^{*} \phi_{k}''^{*}
  \biggr]
  \; ,
\nonumber
\end{aligned}
\end{equation}
where the notation~(\ref{eq:number-shift_multimode}) for the indices of the density matrix has been simplified for these transitional formulas according to the replacements
$
\boldsymbol{ \left\{  \right. } n  \boldsymbol{\left. \right\}}
  \equiv
 \left\{  \right.  
 n_{1}, n_{2}, \ldots , n_{j } , \ldots \boldsymbol{\left. \right\}  }
 \rightarrow n$ and 
 $
\boldsymbol{ \left\{  \right. } n  \boldsymbol{\left. \right\}}_{n_j + q}  
  \equiv
 \left\{  \right.  
 n_{1}, n_{2}, \ldots , n_{j }+ q , \ldots \boldsymbol{\left. \right\}  }
 \rightarrow n_{j} +q$ (with appropriate adjustments);
 for example,
$
 {\rho} \bigl({ \boldsymbol{ \left\{  \right. } n  \boldsymbol{\left. \right\}  } ;
  \boldsymbol{ \left\{  \right. } n  \boldsymbol{\left. \right\}  } } \bigr)
\rightarrow   \rho_{n;n} 
 $ 
 and 
 $
 {\rho} \bigl({ \boldsymbol{ \left\{  \right. } n  \boldsymbol{\left. \right\}  } ;
  \boldsymbol{ \left\{  \right. } n  \boldsymbol{\left. \right\}  }_{n_j +2} } \bigr)
\rightarrow   \rho_{n;n_{j}+2} 
 $, etc.
 In Eq.~(\ref{eq:ME-generic-matrix-elements_multimode}), the required primary identities involving creation and annihilation operators are an appropriate generalization of Eq.~(\ref{eq:ME-primary-matrix-elements}).
 Incidentally, for the last matrix element $\bra{n}  \Phi' \, {\rho} \, \Phi''   \ket{n}$, there is no explicit need to separate diagonal and off-diagonal elements (unlike the asymmetric formulas of the first two matrix elements).

Then, from Eqs.~(\ref{eq:masterequation2}) and (\ref{eq:ME-generic-matrix-elements_multimode}),
the multimode master equation for the reduced field density matrix reads
\begin{equation}
\begin{aligned}
&
\dot{\rho}_{\rm diag} (  \boldsymbol{ \left\{  \right. } n  \boldsymbol{\left. \right\}  } ) 
 =
    - 
     \sum_{j}
     \left\{
     R_{{\mathrm e},\, j}  \big[(n_j+1) \,
   {\rho}_{\rm diag} (  \boldsymbol{ \left\{  \right. } n  \boldsymbol{\left. \right\}  } )
      - n_j \,
   {\rho}_{\rm diag} (  \boldsymbol{ \left\{  \right. } n  \boldsymbol{\left. \right\}  }_{n_j -1} )
            \big] \right.
          \\
           &
          \qquad  
            \left.
            +  
        R_{{\mathrm a},\, j} \big[ n_j \,
    {\rho}_{\rm diag} (  \boldsymbol{ \left\{  \right. } n  \boldsymbol{\left. \right\}  } )
      - (n_j + 1)  \,
     {\rho}_{\rm diag} (  \boldsymbol{ \left\{  \right. } n  \boldsymbol{\left. \right\}  }_{n_j +1} )
                           \big] \right.
                           \\
    &
            \qquad  
   -
   \left[
    S_{+, j } 
     \sqrt{(n_{j}+1)(n_{j}+2)} \, 
     {\rho} \bigl({ \boldsymbol{ \left\{  \right. } n  \boldsymbol{\left. \right\}  }_{n_j +2}  ;
  \boldsymbol{ \left\{  \right. } n  \boldsymbol{\left. \right\}  } } \bigr)
    + S_{-, j }  \sqrt{n_{j}(n_{j}-1)} \,  
 {\rho} \bigl({ \boldsymbol{ \left\{  \right. } n  \boldsymbol{\left. \right\}  } ;
  \boldsymbol{ \left\{  \right. } n  \boldsymbol{\left. \right\}  }_{n_j -2} } \bigr)
    \right.
    \\
    &
            \qquad  
   \left.
    \left.  
    - \left( S_{+, j }  + S_{-, j } \right)  \sqrt{n_{j}(n_{j}+1)} \,
    {\rho} \bigl({ \boldsymbol{ \left\{  \right. } n  \boldsymbol{\left. \right\}  }_{n_{j} +1 } ;
  \boldsymbol{ \left\{  \right. } n  \boldsymbol{\left. \right\}  }_{n_{j} -1} } \bigr)
    +  \mathrm{h.c.} 
\right]         
\right\}
\\
&
\qquad
-
\sum_{j \neq k} \sum_{\epsilon', \epsilon'' =  \pm} 
\left\{
\left[ R^{(+)}_{ (\epsilon', \epsilon'' ); j,k} \;
s^{(+)}_{ (\epsilon', \epsilon'' ); n_{j}, n_{k}} \;
 {\rho} \bigl( { \boldsymbol{ \left\{  \right. } n  \boldsymbol{\left. \right\}  }_{n_{j} -\epsilon' ,n_{k} - \epsilon''}} ;
  \boldsymbol{ \left\{  \right. } n  \boldsymbol{\left. \right\}  }
   \bigr)
    +  \mathrm{h.c.} \right]
\right.
\\
&
\qquad  \qquad \qquad 
-
\left.
  \left[ R^{(-)}_{ (\epsilon', \epsilon'' ); j,k} \;
s^{(-)}_{ (\epsilon', \epsilon'' ); n_{j}, n_{k}} \;
 {\rho} \bigl({ \boldsymbol{ \left\{  \right. } n  \boldsymbol{\left. \right\}  }_{n_{j} -\epsilon' } ;
  \boldsymbol{ \left\{  \right. } n  \boldsymbol{\left. \right\}  }_{n_{k} + \epsilon''} } \bigr)
    +  \mathrm{h.c.} \right]
    \right\}
    \; ,
\end{aligned}
 \label{eq:master_equation_multimode-offdiagonal}
\end{equation}
where the factors
$s^{(\pm)}_{ (\epsilon', \epsilon'' ); n_{j}, n_{k} } =
\sqrt{ \left[n_{j}+\frac{1}{2}(1- \epsilon')\right]  \left[n_{k}+\frac{1}{2}(1 \mp \epsilon'')\right]}$
arise from the creation/annihilation operator normalization (and $\epsilon = \pm 1$, when used as an algebraic factor as opposed to simple index). In addition, in Eq.~(\ref{eq:master_equation_multimode-offdiagonal}), 
the emission ($ R_{{\mathrm e},\, j} $),
 absorption ($ R_{{\mathrm a},\, j} $), and squeezing coefficients
($S_{\pm, j }$), for each mode $j$, are defined as before [Eqs.~(\ref{eq:P_ex_expression})--(\ref{eq:P_ab_expression}),(\ref{eq:P_ex_explicit})--(\ref{eq:P_ab_explicit}), and \ref{eq:squeezing-coefficients})]. 
The new coefficients
\begin{equation}
\begin{aligned}
R^{(\pm)}_{ (\epsilon', \epsilon'') ; j,k} 
& =
 \mathfrak{r} \, g^2 
 {\mathcal J}^{(\pm)}_{ (\epsilon', \epsilon'') ; j,k} 
\\
{\mathcal J}^{(\pm)}_{ (\epsilon', \epsilon'') ; j,k}  & =
 \int_{\tau_i}^{\tau_f} d\tau'\, 
 \int_{\tau_i}^{\tau'} d\tau'' 
 \, e^{\mp i\nu\tau'}  e^{\pm i\nu\tau''}
 \phi_{(\epsilon'); j} (\mathbf{r} (\tau'),t(\tau')) \, \phi_{(\epsilon'');k} (\mathbf{r} (\tau''),t(\tau''))
 \; ,
 \label{eq:mode-correlation-coefficients}
 \end{aligned}
\end{equation}
which generalize Eq.~(\ref{eq:integral-structural}), measure the degree of correlation of pairs of modes in all possible combinations; in particular, they satisfy the conjugate relations 
$\left[ \mathcal{J}^{(\pm)}_{(\epsilon',\epsilon'')}  \right]^{*} =
 \mathcal{J}^{(\mp)}_{(-\epsilon',-\epsilon'')} $, which can be applied to Eq.~(\ref{eq:master_equation_multimode-offdiagonal}).
 
Remarkably, Eq.~(\ref{eq:mode-correlation-coefficients}) is more general than stated above, as it can be extended to the case $j=k$ as well. In effect, Eq.~(\ref{eq:master_equation_multimode-offdiagonal}) can be succinctly written in a more compact form as an unrestricted sum $\sum_{j, k} \sum_{\epsilon', \epsilon'' =  \pm} $, for all $j,k$ that includes the diagonal elements---but the formulas for the diagonal normalization coefficients need to be adjusted. This general validity of Eq.~(\ref{eq:mode-correlation-coefficients}) will be used in 
Appendix~\ref{app:injection-averages_diagonality}, for the injection averages.

Finally, it is also possible to extend Eq.~(\ref{eq:master_equation_multimode-offdiagonal}) to give the off-diagonal matrix elements of $\dot{\rho}$ on the left-hand side of a similar equation. As these off-diagonal elements are not of direct use for the present work, we omit their expressions, which otherwise give obvious extensions of the right-hand side.

\section{Coarse-grained injection averages and density-matrix diagonality}
 \label{app:injection-averages_diagonality}
In this appendix, we address the properties of the injection-averaging procedure, and outline
the proof of the diagonal-reduction property for random injection times, leading to Eq.~(\ref{eq:master_equation_final_multimode}). 

The general results for the field master equation, as derived in Appendix~\ref{app:master_equation_derivation}, apply to the reduced density matrix $\rs{\mathcal P}$ obtained by tracing out the atomic degrees of freedom, but without enforcing the coarse-graining averaging procedure. Such ``microscopic'' or ``bare'' results display the structural form of the master equation~(\ref{eq:master_equation_multimode-offdiagonal}), namely, the correct pattern of matrix elements, but do not give the macroscopic coefficients that correspond to an experimental setup with a specific injection-time distribution for the atomic cloud as reservoir. The coefficients that complete the characterization of Eq.~(\ref{eq:master_equation_multimode-offdiagonal}) are $R^{(\pm)}_{ (\epsilon', \epsilon'') ; j,k} $, 
given in Eq.~(\ref{eq:injection-average_R-coeffs})
(which also describes the mode-diagonal elements).
The outcome of the injection-averaging procedure at the level of the master equation is to make the replacement 
$R^{(\pm)}_{ (\epsilon', \epsilon'') ; j,k} \longrightarrow
\overline{ R^{(\pm)}_{ (\epsilon', \epsilon'') ; j,k} }$.
In particular, this procedure may generate cancellations and additional symmetries, depending on the details of the injection process.

We will now proceed with the specific averaging for the cloud falling into a generalized Schwarzschild black hole. The field modes, in their interaction with an individual atom, have a functional form $ \phi_{\boldsymbol{s}} ({\bf r} (\tau), t(\tau))$, 
which includes a 
dependence on the injection time $t_{i}$. From the general theory leading to the microscopic coefficients, one can introduce this distribution of initial times by the replacement
 $t_{i} \rightarrow t_{0} + \Delta t_{i}$, where $t_{0}$ is a fiducial initial time parameter, and then perform the average, according to the operational rule of Eq.~(\ref{eq:injection-average}), on any relevant field quantity $X^{\mathcal P}$.
 Without any loss of generality, the initial value $t_{0}=0$ can be chosen, so that $\Delta t_{i} = t_{i} $;
 then,
  \begin{equation}
\overline{X^{\mathcal P}}
=
 \int d t_{i} \,  f ( t_{i}) 
X^{\mathcal P} (t_{i})
\; .
\label{eq:injection-average_2}
\end{equation}
In Eq.~(\ref{eq:master_equation_multimode-offdiagonal}), this average behavior is encoded in the coefficients
$R^{(\pm)}_{ (\epsilon', \epsilon'') ; j,k} 
 =
 \mathfrak{r} \, g^2 
 {\mathcal J}^{(\pm)}_{ (\epsilon', \epsilon'') ; j,k} $, so that, from Eq.~(\ref{eq:mode-correlation-coefficients}),
\begin{equation}
\begin{aligned}
R^{(\pm)}_{ (\epsilon', \epsilon'') ; j,k}
&
=
 \int d t_{i} \, 
R^{(\pm)}_{ (\epsilon', \epsilon'') ; j,k}
 f ( t_{i}) 
\\
&
=
 \mathfrak{r} \, g^2 
 \;
 \int d t_{i} \, 
 \int\displaylimits_{I }  
d^{2} \tau
 \, e^{\mp i\nu\tau'}  e^{\pm i\nu\tau''}
 \phi'_{(\epsilon'); j}  \, \phi''_{(\epsilon'');k} 
 \biggr|_{\substack{t' \rightarrow t' + t_{i}  
 \\ t'' \rightarrow t'' + t_{i} }}
  f ( t_{i}) 
\; ,
\label{eq:injection-distribution_R-coeffs}
\end{aligned}
\end{equation}
with the notation $ \phi'_{(-); j} =\phi_{{\boldsymbol{s_{j}}}} (\mathbf{r} (\tau'),t(\tau'))$, 
$ \phi''_{(+); j} =\phi^{*}_{{\boldsymbol{s_{j}}}} (\mathbf{r} (\tau''),t(\tau''))$, etc. 
As displayed in Eq.~(\ref{eq:injection-distribution_R-coeffs}), the initial-time shift with a statistical variable $t_{i}$ can be directly enforced inside the integral, with the replacements 
$t' \rightarrow t' + t_{i}$,  $t'' \rightarrow t'' + t_{i}$.
Then, the functional form of the integrand in Eq.~(\ref{eq:injection-average_R-coeffs}) implies that these shifts 
only affect the product $ \phi'_{(\epsilon'); j}  \, \phi''_{(\epsilon'');k} $, which is thus solely responsible for the $t_{i} $ dependence. 

An important point in the evaluation of the integrals in
Eqs.~(\ref{eq:injection-average_2}) and (\ref{eq:injection-distribution_R-coeffs}) 
(and similar relations) is the choice of integration limits for the formal variable $t_{i}  $.
These can be effectively pushed to infinity for a random distribution: $ \int d t_{i} \equiv   \int_{-\infty}^{\infty} d t_{i} $ due to the short memory of the reservoir in the Markovian approximation.
More precisely, there are several time scales involved:
the cloud memory time $\Delta t_{M}$,
the atomic time scale $\Delta t_{\mathcal{A}} \sim 1/\nu$,
the field characteristic time scale $ \Delta t_{\mathcal{P}} \sim 1/\omega \sim 2 \pi/\kappa$ (from the dominant part of the Planck distribution), and the atomic cloud injection scale 
 $ \Delta t_{C} \sim T $ (which is used for the averaging integrals).
There is an associated hierarchy of time scales:
$\Delta t_{M} , \Delta t_{\mathcal{A}} \ll \Delta t_{\mathcal{P}} \ll \Delta t_{C} $, where the right-most side of the inequality gives the condition that the integration interval $T$ needs to satisfy for the establishment of a sufficiently random distribution.

Therefore, under such conditions, with the replacements 
$t' \rightarrow t' + t_{i}$,  $t' \rightarrow t'' + t_{i}$,
 Eq.~(\ref{eq:injection-distribution_R-coeffs}) leads to
 \begin{equation}
\overline{ R^{(\pm)}_{ (\epsilon', \epsilon'') ; j,k}}
=
\left[
 \int d t_{i} \;
e^{
  i  \left(  \omega_{j} \epsilon' +  \omega_{k} \epsilon''  \right) \, t_{i} } \,  f ( t_{i}) 
  \right]
   \;
 \left. R^{(\pm)}_{ (\epsilon', \epsilon'') ; j,k}  \right|_{\text{microscopic}} 
 \; ,
 \label{eq:injection-average_R-coeffs}
 \end{equation}
where
the
coordinate-time dependence of the modes is explicitly given by 
$ \phi_{\boldsymbol{s}} ({\bf r} , t )  =
\varphi_{\boldsymbol{s}} ({\bf r} ) \, e^{-i \omega  t}$,
whence 
$ \phi_{(\epsilon); j} =\varphi_{(\epsilon); j} \, e^{i \epsilon \omega t}$. 
As can be seen from Eq.~(\ref{eq:injection-average_R-coeffs}),
the microscopic coefficient gets modified by an integral prefactor that is governed by an oscillatory integrand inherited from the time dependence $ \phi_{\boldsymbol{s}} ({\bf r} , t ) e^{-i \omega  t}$ 
of the modes themselves.

In this paper, we are interested in an atomic cloud with a statistical distribution of {\it random injection times\/}. Then, in Eq.~(\ref{eq:injection-average_R-coeffs}), we can choose a uniform distribution 
$ f ( t_{i}) = 1/T$; 
thus, the randomness inherent in $t_{i}$ will make the integral prefactor 
in Eq.~(\ref{eq:injection-average_R-coeffs}),
average out to zero unless the exponent itself is zero.
Specifically, the critical condition arises from
 \begin{equation}
 \lim_{T \rightarrow \infty} 
 \frac{1}{T} \int_{-T/2}^{T/2}  d t_{i} \; e^{i \gamma t_{i} }
=\delta_{\gamma, 0}
  \label{eq:diagonality-integral}
\; ,
 \end{equation}

where the Kronecker delta is zero unless $\gamma = 0$; here,
$ \gamma = \omega_{j} \epsilon' +  \omega_{k} \epsilon''  $. 
Thus, the necessary condition for 
$\overline{ R^{(\pm)}_{ (\epsilon', \epsilon'') ; j,k}}$
to be nonvanishing is 
 \begin{equation}
  \omega_{j} \epsilon' +  \omega_{k} \epsilon''  = 0
  \label{eq:diagonality-constraint}
\; .
 \end{equation}
As the frequencies $  \omega_{j} $ and $\omega_{k} $ are both positive, the condition~(\ref{eq:diagonality-constraint}) can only be satisfied when $\epsilon'$ and $\epsilon''$ have opposite signs; this implies that 
 \begin{equation}
  \epsilon' = -  \epsilon''
  \; \; \; \text{and} \; \; \; 
    \omega_{j} =\omega_{k}
   \label{eq:diagonality-condition}
\; .
 \end{equation}
Consequently, the random-time injection-averaged coefficients are zero unless the mode functions are:
(i) complex conjugates of each other ($ \epsilon' = -  \epsilon''$); and 
(ii) they correspond to the same frequency $\omega_{j} =\omega_{k}$. The former condition is also equivalent to the appearance of pairs of creation and annihilation operators in the matrix elements, thus corresponding to diagonal elements of the field density matrix
when the frequency identity is enforced.
This argument shows that Eq.~(\ref{eq:diagonality-condition}) is a diagonality condition implied by a random distribution of injection times. Under such conditions, the relevant coarse-grained field density matrix is diagonal and given by Eq.~(\ref{eq:master_equation_final_multimode}). Specifically, Eq.~(\ref{eq:diagonality-condition}) shows that all the single-mode squeezing coefficients~(\ref{eq:squeezing-coefficients}) and mode-off-diagonal coefficients~(\ref{eq:mode-correlation-coefficients}) become zero when the averaging procedure is enforced. By contrast, the emission and absorption rates $R_{{\mathrm e},j}$ and $R_{{\mathrm a},j}$ are allowed to be nonzero, as Eq.~(\ref{eq:diagonality-condition}) is valid for them----this can be seen by comparison with the general expressions of Eq.~(\ref{eq:Pe-Pa_from_J-integrals}). In addition, direct inspection of Eq.~(\ref{eq:master_equation_multimode-offdiagonal}) verifies that the only nonvanishing terms in the master equation are diagonal.

\section{Geodesic equations in generalized Schwarzschild metric}
\label{app:geodesics-Schwarzschild}

 Any static and spherically symmetric metric defined by Eq.~(\ref{eq:RN_metric}),
 has invariance under time translations and under spatial rotations involving $(D-1)(D-2)/2$ planes. All of the angular momentum components but one can be fixed to define a single plane for the orbit where an azimuthal angle $\phi$ can be used. 
The corresponding Killing vectors are $  \boldsymbol\xi = \partial_{t}$ and $\boldsymbol\eta = \partial_{\phi}$.

The free-fall spacetime trajectories are the geodesics. Their first-order form for timelike geodesics can be easily
set up with the constants of the motion: the mass $\mu$  and the conserved quantities
$
e = - \boldsymbol\xi \cdot {\bf u} = E/\mu$ (energy per unit mass)
and $\ell = \boldsymbol\eta \cdot {\bf u} =L/\mu$ (angular momentum per unit mass), where $ {\bf u}$
is the spacetime velocity. 
Thus, the geodesic equations are
\begin{align}
    \frac{dt}{d\tau} &= \frac{e}{f(r)}\;.\label{eq:geodesic_t} \\
      \frac{dr}{d\tau} &= -\sqrt{e^2- f(r)\lrp{1+\frac{\ell^2}{r^2}}}\;,\label{eq:geodesic_tau}\\
   \frac{d\phi}{d\tau}  & = \frac{\ell}{ r^2} \; \label{eq:geodesic_phi}
   \; .
\end{align}
In Eq.~(\ref{eq:geodesic_tau}), it is assumed the in-falling motion of the atom is analyzed. 

The geodesic equations can be integrated to derive the atom's proper time $\tau$ and the Schwarzschild coordinate time $t$ in terms of the radial variable $r$, 
\begin{align}
    \tau - \tau_{i} &= - \int_{r_i}^{r} \frac{dr}{\sqrt{e^2- f(r)\lrp{1+ {\ell^2}/{r^2}}}} 
    \equiv F(r)
      \label{eq:tau_integration} \\
    t  - t_{i} &= -\int_{r_i}^{r} dr \, \frac{e/f}{\sqrt{e^2- f(r)\lrp{1+ {\ell^2}/{r^2}}}} 
    \equiv G(r)
     \; , \label{eq:t_integration}
\end{align}
where $r_i$ is the radial coordinate of a fiducial point---for the cloud of atoms this a point, or radial coordinate value, from which the atoms are injected with initial specific energy $e$ and initial specific angular momentum $\ell$.
 Equations~(\ref{eq:tau_integration}) and (\ref{eq:t_integration})
 have the functional forms 
$ \tau - \tau_{i} =  F(r)$ and
$ t- t_{i} =  G(r)$, so that 
\begin{equation}
 \tau- \tau_{i} =  H (t-t_{i} ) 
 \; ,  \; \text{where} \; \; \; 
 H=F \circ G^{-1}
 \label{eq:tau-from-t}
 \end{equation}
  is the composite function that implies a correspondence generically of the form $\tau=\tau(t)$,
with specific initial values $\tau_{i}$ and $t_{i}$. 

The near-horizon expansions of Eqs.~(\ref{eq:tau_integration}) and (\ref{eq:t_integration}),
up to first order in $x=r-r_+$, are
\begin{align}
    \tau &= -\frac{x}{e}+\mathrm{ const.} +  \mathcal{O}(x^2)\;,\label{eq:tau_in_x_app} \\
    t &= -\frac{1}{f_+'}\ln x - C x
    + \mathrm{ const.} + \mathcal{O}(x^2)\;, \label{eq:t_in_x_app} 
\end{align}
where 
\begin{equation}
C = \frac{1}{2}\lrb{\frac{1}{e^2}\lrp{1+\frac{\ell^2}{r_+^2}}-\frac{f_+''}{(f_+')^2}} 
\; 
\label{eq:constant-C}
\end{equation}
is a constant that depends on the conserved quantities $e$ and $\ell$, as well as the black hole parameters
$r_{+}$ and $f_{+}'$ carried by $f(r)$.
Equations~(\ref{eq:tau_in_x_app}) and (\ref{eq:t_in_x_app}) have the correct behavior in the neighborhood of the horizon, with a logarithmically divergent coordinate time but finite proper time.

\end{appendix}



\begin{thebibliography}{}
\bibitem{hawking76}
S. W. Hawking, Black holes and thermodynamics, Physical Review D, 13(2), 191 (1976).

\bibitem{BH-thermo_reviews}
R.~M. Wald, 
The thermodynamics of black holes, Living Rev. Rel. 4, 6 (2001);
and references therein.

\bibitem{bekenstein1972}
J. D. Bekenstein, Black holes and the second law. Lettere al Nuovo Cimento, 4, 737-740 (1972).

\bibitem{bekenstein1973}
J. D. Bekenstein, Black holes and entropy, Physical Review D, 7(8), 2333 (1973);

\bibitem{bekenstein1974}
J. D. Bekenstein, Generalized second law of thermodynamics in black-hole physics, Physical Review D, 9(12), 3292 (1974).

\bibitem{christodoulou}
D. Christodoulou, Reversible and irreversible transformations in black-hole physics. Physical Review Letters, 25, 1596 (1970).

\bibitem{hawking71}
S. W. Hawking, Gravitational radiation from colliding black holes. Physical Review Letters, 26, 1344 (1971).

\bibitem{hawking72}
S. W. Hawking, Black holes in general relativity. Communications in Mathematical Physics, 25, 152-166 (1972).

\bibitem{bardeen-carter-hawking1973}
J. M. Bardeen, B. Carter, and S. W. Hawking, The four laws of black hole
mechanics. Communications in Mathematical Physics 31, 161 (1973).

\bibitem{hawking74}
S. W. Hawking, Black hole explosions? Nature, 248(5443), 30-31 (1974).

\bibitem{hawking75}
S. W. Hawking, Particle creation by black holes, Communications in Mathematical Physics, 43(3), 199-220 (1975).

\bibitem{unruh76}
W. G. Unruh, Notes on black-hole evaporation, Physical Review D, 14(4), 870 (1976).

\bibitem{fulling76}
P. C. Davies, S. A. Fulling, and W. G. Unruh, Energy-momentum tensor near an evaporating black hole, Physical Review D, 13(10), 2720 (1976).

\bibitem{davies77}
P. C. Davies, S. A. Fulling, Radiation from moving mirrors and from black holes, Proceedings of the Royal Society of London. A. Mathematical and Physical Sciences, 356(1685), 237-257 (1977).

\bibitem{scully2018}
M. O. Scully, S. Fulling, D. M. Lee, D. N. Page, W. P. Schleich, A. A. Svidzinsky (2018). Quantum optics approach to radiation from atoms falling into a black hole, Proceedings of the National Academy of Sciences, {\bf 115}(32), 8131, 2018.

\bibitem{camblong2005}
H. E. Camblong, and C. R. Ord\'o\~nez, Black hole thermodynamics from near-horizon conformal quantum mechanics, Physical Review D, {\bf 71}, 104029 (2005).

\bibitem{nhcamblong-sc}
H. E. Camblong and C. R. Ord\'o\~nez, Semiclassical methods in curved spacetime and black hole thermodynamics, Phys. Rev. D, 71, 124040 (2005).

\bibitem{camblong2013}
H. E. Camblong, and C. R. Ord\'o\~nez, Conformal tightness of holographic scaling in black hole thermodynamics, Classical and Quantum Gravity, {\bf 30}, 175007 (2013).

\bibitem{camblong2020}
H. E. Camblong, A. Chakraborty, and C. R.  Ord\`o\~nez, Near-horizon aspects of acceleration radiation by free fall of an atom into a black hole, Phys. Rev. D, 102, 085010, 2020.

\bibitem{azizi2021}
A. Azizi, H. E. Camblong, A. Chakraborty, C. R. Ord\`o\~nez, and M. O. Scully, Acceleration radiation of an atom freely falling into a Kerr black hole and near-horizon conformal quantum mechanics. arXiv preprint arXiv:2011.08368, 2020.

\bibitem{gupta01}
D. Birmingham, K. S. Gupta, and S. Sen, Near-horizon conformal structure of black holes, Physics Letters B, 505(1-4), 191-196 (2001).

\bibitem{vaidya00}
T. R. Govindarajan, V. Suneeta, S. Vaidya, Horizon States for AdS Black Holes. Nucl. Phys. B583, 291-303 (2000).

\bibitem{moretti02}
V. Moretti, and N. Pinamonti, Aspects of hidden and manifest SL(2, R) symmetry in 2D near-horizon black-hole backgrounds, Nuclear Physics B, 647(1-2), 131-152 (2002).

\bibitem{DFF}
V. de Alfaro, S. Fubini and G. Furlan, Conformal Invariance In Quantum Mechanics, Nuovo Cim. A34 569 (1976).

\bibitem{HBAR_part-II}
H. E. Camblong, A. Chakraborty, and C. R.  Ord\`o\~nez, 
Quantum optics meets black hole thermodynamics via conformal quantum mechanics:
II. Thermodynamics of acceleration radiation.

\bibitem{wilson18}
S. A. Fulling, and J. H. Wilson, The equivalence principle at work in radiation from unaccelerated atoms and mirrors, Physica Scripta, 94(1), 014004 (2018).

\bibitem{scullyreview}
J. S. Ben-Benjamin et al., Unruh Acceleration Radiation Revisited, International Journal of Modern Physics A, 34(28), 1941005 (2019).

\bibitem{scully2003}
M. O. Scully, V. V. Kocharovsky, A. Belyanin, E. Fry, and F. Capasso,
Enhancing acceleration radiation from ground state-atoms via cavity quantum electrodynamics, Phys. Rev. Lett., 91, 243004 (2003).

\bibitem{belyanin2006}
A. Belyanin, V. V. Kocharovsky, F. Capasso, E. Fry, M. S. Zubairy, and M. O. Scully, 
Quantum electrodynamics of accelerated atoms in free space and in cavities, Phys. Rev. A, 74, 023807 (2006).

\bibitem{scullybook}
M. O. Scully and M. S. Zubairy, Quantum optics, Cambridge University Press, New York, 1997.

\bibitem{meystrebook}
P. Meystre, and M. Sargent, Elements of quantum optics, Springer Science \& Business Media, 2007.

\bibitem{Birrell-Davies}
N. D. Birrell and P. C. W. Davies, 
Quantum fields in curved space (Cambridge University Press, Cambridge, 1982).


\end{thebibliography}
\end{document}